# Conductivity of $CuO_3$-Chains: Disorder versus Electron-Phonon Coupling

R. Fehrenbacher

*Theoretische Physik, ETH-Hönggerberg, CH-8093 Zürich, Switzerland*[1]

**Abstract**. The optical conductivity of the $CuO_3$-chains, a subsystem of the 1-2-3 materials, is dominated by a broad peak in the mid-infrared ($\omega \approx 0.2$eV), and a slowly falling high-frequency tail. The 1D $t$-$J$-model is proposed as the relevant low-energy Hamiltonian describing the intrinsic electronic structure of the $CuO_3$-chains. However, due to charge-spin decoupling, this model alone cannot reproduce the observed $\sigma(\omega)$. We consider two additional scattering mechanisms: (i) Disregarding the not so crucial spin degrees of freedom, the inclusion of strong potential disorder yields excellent agreement with experiment, but suffers from the unreasonable value of the disorder strength necessary for the fit. (ii) Moderately strong polaronic electron-phonon coupling to the mode involving Cu(1)-O(4) stretching, can be modeled within a 1D Holstein Hamiltonian of spinless fermions. Using a variational approximation for the phonon Hilbert space, we diagonalize the Hamiltonian exactly on finite lattices. As a result of the experimental hole density $\approx 1/2$, the chains can exhibit strong charge-density-wave (CDW) correlations, driven by phonon-mediated polaron-polaron interactions. In the vicinity of half filling, charge motion is identified as arising from moving domain walls, *i. e.*, defects in the CDW. Incorporating the effect of vacancy disorder by choosing open boundary conditions, good agreement with the experimental spectra is found. In particular, a high-frequency tail arises as a consequence of the polaron-polaron interactions.

*PACS Numbers: 71.38, 71.45.L, 74.25, 74.72, 78.30, 78.30.L*

[1]New Address: Science and Technology Center for Superconductivity, Materials Science Division, Argonne National Laboratory, 9700 South Cass Avenue, Argonne, Illinois 60439

# I. Introduction

The 1-2-3 family of high-$T_c$ oxides with its most prominent member $YBa_2Cu_3O_{7-\delta}$ is the only example of the new superconductors which in addition to the common $CuO_2$ planes also contains one-dimensional $CuO_3$-chains contributing to the electronic density of states at the Fermi level.[1] At the same time, this class of compounds is the best characterized among the high-$T_c$ materials, and furthermore, many of the technological applications are based on $YBa_2Cu_3O_{7-\delta}$. However, in the interpretation of many experimental data, it is not obvious how the features of the chains can be separated from those of the planes. The latter are known to drive the superconductivity in *all* high-$T_c$ compounds, and are therefore of primary interest. Only recently, with the availability of single-domain crystals, a clearer picture of the interplay between chains and planes has emerged. In any case, for a complete understanding of the properties of the 1-2-3 compounds, it is absolutely necessary to distinguish, and where possible isolate the features of the $CuO_3$-chains.

One of the physical quantities which contains valuable information about the electronic structure of a system is the optical conductivity $\sigma(\omega)$. Its frequency dependence is determined by both, the character of the ground state, as well as (in principle) *all* the excited states. Experimentally, $\sigma(\omega)$ can be determined quite easily, *e. g.*, from reflectivity measurements by applying a Kramers-Kronig transformation.

The synthesis of detwinned $YBa_2Cu_3O_{7-\delta}$, and $PrBa_2Cu_3O_{7-\delta}$ crystals made it possible to extract the anisotropy of $\sigma(\omega)$ in the crystallographic **a**- (perpendicular), and **b**-direction (parallel to the chains), by polarizing the electric field **E** in one or the other direction. This revealed an interesting and important property of these compounds, namely that $\sigma_b(\omega)$ (with $\mathbf{E} \parallel \mathbf{b}$) shows a strongly enhanced spectral weight compared to $\sigma_a(\omega)$ especially in the infrared region ($\omega \lesssim 1.5$eV). A procedure which suggests itself is then to subtract $\sigma_a(\omega)$ from $\sigma_b(\omega)$, and to associate the corresponding oscillator strength with the $CuO_3$-chains, the main source of anisotropy in these materials.

In all published data (Ref.[1] for $PrBa_2Cu_3O_{7-\delta}$, and Ref.[2, 3, 4] for $YBa_2Cu_3O_{7-\delta}$), these difference spectra $\sigma_{\mathrm{ch}}(\omega) := \sigma_b(\omega) - \sigma_a(\omega)$ show a characteristic broad *mid-infrared peak* centered around 0.2eV, and a *slowly falling high-frequency tail* extending beyond 1eV. This tail falls off much slower than the simple Drude formula ($\sigma \sim 1/\omega^2$ at high frequency) would suggest. At energies below 0.2eV, there seems to be a difference between the two compounds: while in $YBa_2Cu_3O_{7-\delta}$, the conductivity joins onto a Drude-peak above $T_c$,[1] in $PrBa_2Cu_3O_{7-\delta}$ it drops sharply to zero, reflecting the insulating nature of the latter compound. Note, that in the Pr-compound, this behaviour of the conductivity has important implications for possible mechanisms explaining the observed $T_c$-suppression[6]. The above described shape of $\sigma_{\rm ch}(\omega)$ clearly indicates that the electronic structure of the chains is non-trivial, and it stimulated our interest to find a possible explanation.

[1] In fact, this statement is not quite correct. There exists a slight variation of these materials, the so-called 1-2-4 compounds, with its prototype $YBa_2Cu_4O_8$. They contain one-dimensional $CuO_3$ *double-chains*, which also provide states at the Fermi level.

In this article, we offer *two* likely mechanisms which could lead to the observed $\sigma_{\rm ch}(\omega)$: (i) the presence of very strong potential disorder, and (ii) moderately strong polaronic electron-phonon (e-ph) interactions, possibly combined with the effect of vacancy disorder. While the first explanation gives a better (in fact excellent) fit to the experimental data, the second one seems to be more reasonable from a physical point of view, since, as we shall show, the potential disorder strength needed to obtain agreement with the experimental data is hard to justify.

The plan of this paper is as follows: First we shall identify the one-dimensional (1D) *t*-*J*-model to be a good starting point for the description of the electronic structure of the $CuO_3$-chains. However, as is well known by now from various studies, the optical conductivity of the 1D *t*-*J*-model is more or less identical to that of spinless fermions (as long as $J$ is below the threshold for phase separation) because of charge-spin separation. There is barely any spectral weight at finite frequencies [7]. Thus, one is forced to include other interaction terms in the Hamiltonian, which could possibly reproduce the experimentally observed strong scattering.

We continue with a simple extension of the model by including the disorder from oxygen vacancies on the bridging O(1) site. This is expected to be the main source of disorder in the 1-2-3 materials. It is mainly due to the large mobility of the oxygen atoms at higher temperatures. Such vacancies can be modeled by interrupted 1D *t*-*J* chains, but we will show that such a scenario cannot reproduce the observed $\sigma_{\rm ch}(\omega)$. However, as mentioned above, the inclusion of strong potential disorder turns out to yield an excellent fit to the experiment.

As an alternative explanation, we then consider moderately strong e-ph interactions. We suggest that the apical oxygen mode with a displacement in the **c**-direction is likely to be rather strongly coupled to the electrons. Using the $t$-$J$-model as the effective electronic model, we will show that this coupling can be included in the form of a Holstein term, in which the local ionic displacement couples linearly to the local charge. Employing a variational approximation for the phonon Hilbert space, we calculate $\sigma(\omega)$ for finite systems using the Lanczos technique. The effect of the vacancy disorder can be incorporated by the choice of open boundary conditions.

[1] This point is still somewhat controversial. More recent reflectivity measurements of $YBa_2Cu_3O_{7-\delta}$ down to lower frequencies [5], indicated that even above $T_c$ there is no upturn in the conductivity as the dc limit is approached.

Finally, we will discuss the results of our different approaches in the light of the experimental data, and also suggest possible experimental tests of the different mechanisms.

## II. The Electronic Structure of $CuO_3$-chains

When trying to construct a model for the electronic structure of the $CuO_3$-chains, the first thing one should note is the fact that the chain Cu(1) site has the same four-fold oxygen coordination as the planar Cu(2,3) (neglecting the buckling of the O(2,3) atoms out of the plane). Therefore, we can expect a similar Cu-O bonding pattern in both cases. This is confirmed by band-structure calculations for $YBa_2Cu_3O_{7-\delta}$ [8], which apart from the common strongly dispersive planar CuO $pd\sigma$-band also find a chain-band crossing the Fermi energy $E_F$. This band is formed from an antibonding $pd\sigma$ combination of O(1)2$p_y$-Cu(1)3$d_{z^2-y^2}$-O(4)2$p_z$ orbitals, and shows a dispersion similar to the planar band, but only along the chain axis.

Regarding the similarity between planes and chains, it seems reasonable to try a description of the electronic structure by the 1D analog of the Emery model [9]

$$H_{\text{Em}} = \sum_{ijs} \varepsilon_{ij}\, c^\dagger_{is} c_{js} + \sum_i U_i\, c^\dagger_{i\uparrow} c_{i\uparrow} c^\dagger_{i\downarrow} c_{i\downarrow}, \tag{1}$$

where $i, j$ label the Cu(1) and O(1,4) sites, the operators $c^\dagger_{is}$ create 3$d_{x^2-y^2}$-Cu and 2$p_{y,z}$-O holes, and $s$ is the spin index. The $\varepsilon_{ij}$ include the usual on-site energies $\varepsilon_d = 0$ (by convention), $\varepsilon_p$ = 3.6 (which we assume to be the same for O(1,4)), and the nearest neighbour (n.n) hopping integrals $t_{pd} = 1.3$, $t_{pp} = 0.65$ (all values in eV). The cited values correspond to the ones found for the plane [10], which, regarding the above mentioned similarities, and the similar CuO bond lengths, also seem realistic for the chains. Further justification for the use of these values comes from measurements of the optical gap in the compound $Ca_2CuO_3$ [11], in which one finds the same $CuO_3$-chains with a $Cu^{II}$ oxidation state, *i. e.*, in the insulating charge-transfer regime. The observed value of the optical gap is ≈ 2eV, which agrees well with the values found in the *planar* antiferromagnetic charge-transfer insulators like $YBa_2Cu_3O_6$, and therefore suggests a similar value of $\varepsilon_p - \varepsilon_d$.

In the light of this, it seems appropriate to apply the same procedure, as for the $CuO_2$-planes [12], to derive an effective low-energy Hamiltonian from (1), which consists of a *single, strongly correlated* tight-binding band (projected on the subspace of singly occupied sites), and a nearest-neighbour antiferromagnetic exchange interaction, *i. e.*, the 1D *t-J*-model

$$H_{\text{eff}} = -t \sum_{l,s} \mathcal{P}_l\, c^\dagger_{ls} c_{l+1,s} \mathcal{P}_{l+1} + J \sum_{l,s} \left( \mathbf{S}_l \cdot \mathbf{S}_{l+1} - \frac{1}{4} n_l n_{l+1} \right). \tag{2}$$

In this effective description, $l$ labels the Cu(1) sites of the chain, $c_{ls}^{\dagger}$ creates a hole in the Cu $3d_{x^2-y^2}$ orbital with spin $s$, its neighbouring O $2p$-shells being filled ($Cu^{II}$), and $c_{l,s}$ creates a singlet formed from Cu $3d_{x^2-y^2}$ and O $2p_{y,z}$ holes ($Cu^{III}$). The projection operators $\mathcal{P}_l = 1-n_{l\uparrow}n_{l\downarrow}$ eliminate configurations with doubly occupied sites. Again, we argue that the effective parameters $t$ and $J$ should be very similar to the ones derived for the plane [10], and hence, we adopt those values, $t \approx 0.4$eV, $J \approx 0.1$eV.

We use the well-known Kubo formula for the calculation of the optical conductivity $\sigma(\omega)$. In particular, for a one-band tight-binding Hamiltonian like (2), the prescription to calculate $\sigma(\omega)$ at wave vector $\mathbf{q} = 0$ is the following [13]: If we use periodic boundary conditions (PBC), the real (absorbent) part $\sigma_1(\omega)$ of a $t$-$J$ ring can be decomposed into a Drude and a regular term

$$\sigma_1(\omega) = D\delta(\omega) + \sigma_1^{\mathrm{reg}}(\omega), \tag{3}$$

where the Drude weight $D$ is given by

$$D = \pi N \left. \frac{\partial^2 E_0(\Phi)}{\partial \Phi^2} \right|_{\Phi=\Phi_0}, \tag{4}$$

with $E_0(\Phi)$ being the ground state energy of the system in the presence of a time dependent magnetic flux $\Phi$ through the ring, and $N$ is the number of lattice sites. The flux can be generated by a uniform vector potential along the ring (x-direction), $\mathbf{A} = Ae^{-i\omega t}\mathbf{e}_x$ ($\Phi = NAe/(\hbar c)$) which modifies the hopping by the usual Peierls phase factor $t \curvearrowright t\exp(\pm i\Phi/L)$. The lattice constant $a_0$ has been set to 1. $\Phi_0$ is the flux for which the total energy is minimal. Usually either $\Phi_0 = 0$ (corresponding to PBC) or $\Phi_0 = \pi$ [corresponding to antiperiodic boundary conditions, (APBC)], depending on the band filling. For the $t$-$J$-model, *e. g.*, one obtains $\Phi_0 = 0$ if the number of electrons $n_e$ satisfies $n_e = 4m + 2$, but $\Phi_0 = \pi$ for $n_e = 4m$ ($m$ a positive integer). In general, at $\Phi = \Phi_0$, the expectation value $\langle j_x \rangle$ of the current operator $j_x$ at $\mathbf{q} = 0$

$$j_x = it \sum_{l,s} \left( c_{l+1,s}^{\dagger} c_{l,s} - c_{ls}^{\dagger} c_{l+1,s} \right) \tag{5}$$

in the ground state $|\Psi_0(\Phi_0)\rangle$ vanishes, which is required to make (4) applicable. The regular term $\sigma_1^{\mathrm{reg}}(\omega)$ can be written as

$$\sigma_1^{\mathrm{reg}}(\omega) = \frac{\pi e^2}{N} \sum_{m \neq 0} \frac{|\langle \Psi_m | j_x | \Psi_0 \rangle|^2}{E_m - E_0} \, \delta[\omega - (E_m - E_0)], \tag{6}$$

where $\Psi_0, \Psi_m$ are the eigenstates of the Hamiltonian (2) at $\Phi = \Phi_0$. For the total integrated spectral weight $\sigma_{\text{tot}}$ under $\sigma_1(\omega)$, there exists an $f$-sum rule

$$\sigma_{\text{tot}} := \int_0^\infty \sigma_1(\omega) d\omega = -\frac{\pi e^2}{2N} \langle \Psi_0 | T | \Psi_0 \rangle, \tag{7}$$

with $T$ the kinetic energy in (2). With the help of this sum rule it is possible to give an alternative expression for the Drude weight $D$

$$D = \frac{\pi e^2}{N} \left[ -\langle \Psi_0 | T | \Psi_0 \rangle - 2 \sum_{m \neq 0} \frac{|\langle \Psi_m | j_x | \Psi_0 \rangle|^2}{E_m - E_0} \right], \tag{8}$$

which is helpful to check the calculated results.

This formalism has been used to calculate $\sigma(\omega)$ for the 1D $t$-$J$-model [7] numerically on finite chains employing the widely used Lanczos algorithm. The (at first sight) surprising result was that the scattering introduced by the strong correlations is barely noticeable: The integrated spectral weight at finite $\omega$ appears to be an extremely small fraction of the total weight $\sigma_{\text{tot}}$, which indicates that the optical response in the 1D $t$-$J$-model is almost identical to the case of spinless fermions. The deeper reason for this is the separation of spin and charge, which is a well-known and generic feature of 1D models with interacting electrons. These models do not exhibit Fermi-liquid behaviour, but belong to a different universality class, so-called *Luttinger liquids* [14].

For the 1D $t$-$J$-model, an exact solution exists only at the supersymmetric point $J = 2t$, at which the difference in the charge and spin velocity can be rigorously shown [15]. However, in the small $J/t$ limit, we approach the strong coupling limit of the 1D Hubbard model, for which only quite recently Ogata and Shiba have shown [16] that the ground state Bethe-ansatz wave function factorizes into a charge and spin part. The charge part corresponds to the ground state of the corresponding system of spinless fermions, and the spin part to the 1D AF Heisenberg ground state. In fact, the spin-charge decoupling can be intuitively understood from the fact that in 1D, the charge motion *alone* is not capable of changing the spin configurations, since this requires the existence of closed loops in the lattice topology.

This factorization, *i. e.*, the charge-spin separation, has the effect, that in the small $J/t$ limit, the optical conductivity consists basically of the one expected from non-interacting spinless fermions, since the electric field couples only to the charge degrees of freedom. Consequently, $\sigma(\omega)$ is dominated ($\gtrsim$ 99% of the total weight) by a Drude peak, and only very little weight at finite $\omega$. Of course, as we cross the phase separation line for very large values of $J \gtrsim 3t$, the situation changes: (i) The total weight decreases with increasing $J$, and (ii) more and more spectral weight shifts from the Drude peak to finite $\omega$, until we end up with an insulator at $J = \infty$. But these values are far too large to explain the experimentally observed spectrum. We therefore conclude that the 1D $t$-$J$-model alone cannot give a satisfactory explanation for the optical response of the $CuO_3$-chains. So we have to include additional terms in the basic Hamiltonian (2).

## III. Disorder Effects

Various experimental studies of the high-$T_c$ oxides in general, and the 1-2-3 materials in particular have shown that it is very difficult to control the oxidation, *i. e.*, the oxygen content of a sample. The high mobility of the oxygen atoms (in some compounds even down to below room temperature) makes it practically impossible to obtain exactly stochiometric compounds. This is especially severe in the case of the $REBa_2Cu_3O_{7-\delta}$ family, where the O(1) atoms which form the Cu-O bonds along the $CuO_3$-chains, have an extremely high mobility. Therefore, one expects (and finds) a large degree of disorder in the chains, in particular originating from O vacancies, and possibly interstitials.

A first extension of the *t*-*J*-model should therefore be the incorporation of disorder. Since we have seen in the last section that in 1D for realistic values of $J/t \approx 1/4$ the spin exchange scattering hardly alters the optical conductivity as compared to non-interacting spinless fermions (corresponding to $J = 0$), we will neglect the *J*-term in the calculations of this section, *i. e.*, we treat the electrons as spinless. This simplifies the calculations dramatically, because the problem reduces to a single-particle one.

We will consider two kinds of disorder: (i) disorder in the hopping matrix element $t$, and (ii) randomly disordered on-site energies $\varepsilon_l$. Type (i) should be a realistic model for oxygen vacancies (see Fig. 1) taking into account that across a vacant oxygen site the effective overlap matrix element will be drastically reduced. Furthermore, a vacancy can also be expected to produce a strong electrostatic potential arising from the missing charge of the $O^{-2}$ ion. A possible source of *potential disorder* leading to random on-site energies $\varepsilon_l$, are, *e. g.*, oxygen interstitials (see also Fig. 1) or other defects.

The spinless Hamiltonian including the effect of disorder can be written as

$$H_{\rm dis} = \sum_l \varepsilon_l c_l^\dagger c_l - \sum_l t_l (c_{l+1}^\dagger c_l + \text{h.c.}) \quad , \tag{9}$$

where the $t_l, \varepsilon_l$ are random variables with certain probability distributions which will be specified later. This Hamiltonian contains only single-particle terms, and therefore its spectrum as well as the associated single-particle orbitals are easily obtained by numerical diagonalization of the corresponding tridiagonal matrix. For the calculation of $\sigma_1^{\rm reg}(\omega)$, we can still use Eq. (6) with the modified current operator

$$j_x = i \sum_l t_l (c_{l+1}^\dagger c_l - c_l^\dagger c_{l+1}). \tag{10}$$

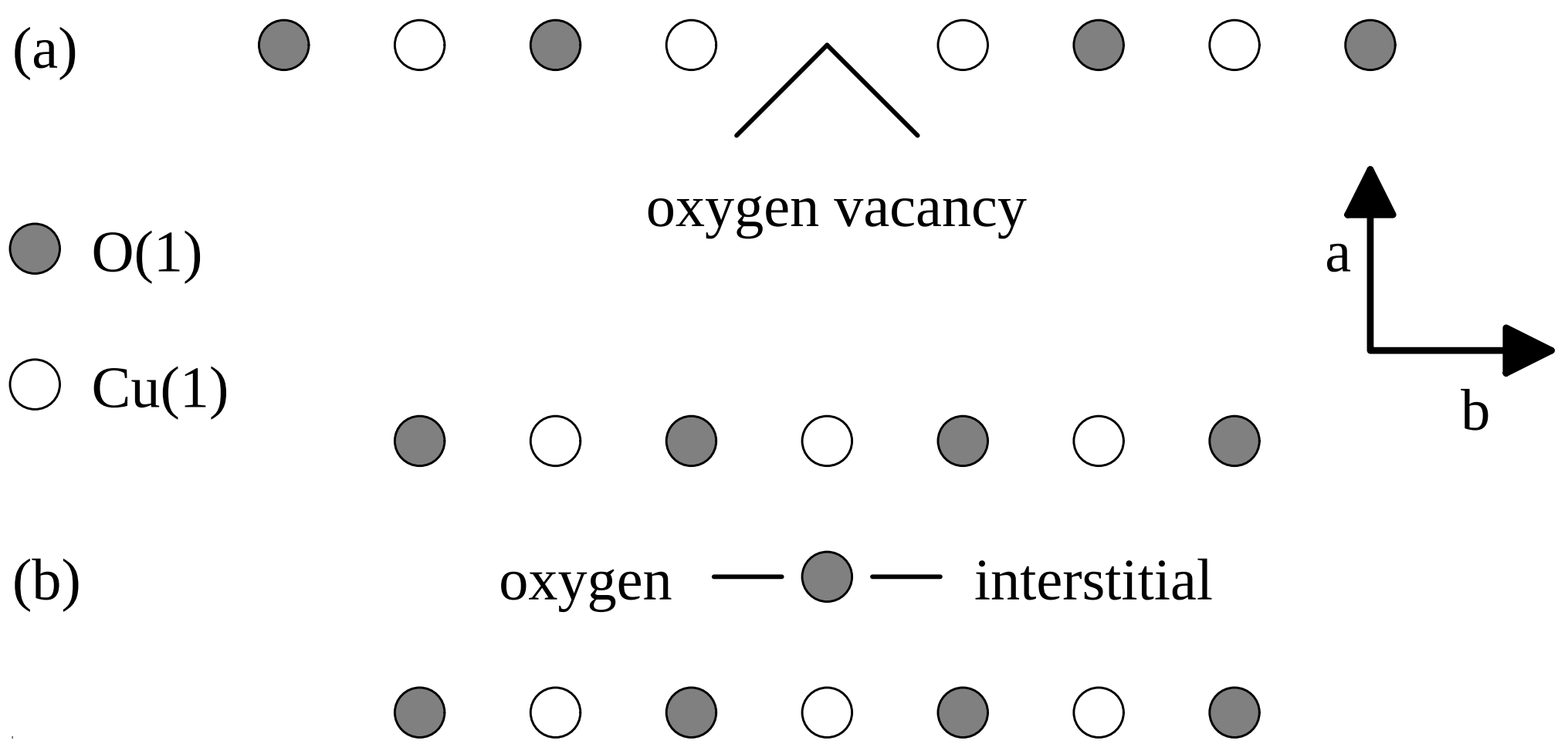

**Figure** 1: Schematic diagram of an oxygen vacancy (a) and an oxygen interstitial (b) in the $CuO_3$-chains.

$H_{\rm dis}$ is diagonal in the single-particle states $\tilde{c}_r$

$$H_{\rm dis} = \sum_r \tilde{\varepsilon}_r\, \tilde{c}_r^\dagger\, \tilde{c}_r$$

$$c_l = \sum_r \alpha_{lr} \tilde{c}_r \qquad c_l^\dagger = \sum_r \alpha_{lr} \tilde{c}_r^\dagger \quad . \tag{11}$$

The coefficients are

$$\alpha_{rl}^* = \alpha_{lr} = \langle \tilde{\phi}_r | \phi_l \rangle$$

$$|\phi_l\rangle = c_l^\dagger |0\rangle \qquad |\tilde{\phi}_r\rangle = \tilde{c}_r^\dagger |0\rangle. \tag{12}$$

The many-particle eigenstates are simple Slater determinants

$$|\Psi(\{n_r\})\rangle = \prod_{\{n_r\}} \tilde{c}_{n_r}^\dagger |0\rangle, \tag{13}$$

where $\{n_r\}$ specifies $n_e$ single particle levels ($n_e$ = number of electrons). If we label the single-particle levels $\tilde{\varepsilon}_r$ in ascending order ($\tilde{\varepsilon}_r \le \tilde{\varepsilon}_{r+1}$), the ground state is given by

$$|\Psi_0\rangle = \prod_{r \le n_e} \tilde{c}_r^\dagger |0\rangle. \tag{14}$$

In order to calculate expectation values of $j_x$, we express $j_x$ in terms of the $\tilde{c}_r$

$$\begin{aligned} j_x &= i \sum_l \left[ t_l \sum_{r,s} (\alpha_{lr}\alpha_{l+1,s} - \alpha_{l+1,r}\alpha_{ls})\, \tilde{c}_r^\dagger\, \tilde{c}_s \right] \\ &= i \sum_{r,s} \tilde{\alpha}_{rs} \tilde{c}_r^\dagger\, \tilde{c}_s, \end{aligned} \tag{15}$$

with

$$\widetilde{\alpha}_{rs} = \sum_l t_l \, (\alpha_{lr}\alpha_{l+1,s} - \alpha_{l+1,r}\alpha_{ls}). \tag{16}$$

With the help of Eq. (15) it becomes very easy to calculate $\sigma(\omega)$ at $T = 0$, since $j_x$ creates only single particle-hole excitations above the ground state, and hence $\sigma_1^{\text{reg}}(\omega)$ can be written as

$$\sigma_1^{\text{reg}}(\omega) = \frac{\pi e^2}{N} \sum_{s \leq n_e, r > n_e} \frac{|\widetilde{\alpha}_{rs}|^2}{\widetilde{\varepsilon}_r - \widetilde{\varepsilon}_s} \, \delta[\omega - (\widetilde{\varepsilon}_r - \widetilde{\varepsilon}_s)]. \tag{17}$$

In all of the calculations of this section, we shall use open boundary conditions. Consequently, the Drude weight *must* vanish (more precisely, it is shifted to a delta peak at a frequency of the order $t/N$), and the entire spectral weight appears at finite frequencies. However, for chain lengths $N \gtrsim 100$, one has essentially reached the thermodynamic limit for the shape of $\sigma_1^{\text{reg}}(\omega)$ at frequencies $\omega \ll t/N$, in which we are interested here.

A realistic model for a situation where the oxygen vacancies are the dominant source of disorder (as seems to be the case for the $CuO_3$-chains) can be constructed by an appropriate choice of the coefficients $\varepsilon_l, t_l$: For a single vacancy between Cu sites $l$ and $l+1$, we assume a strongly reduced hopping matrix element $t'/t \ll 1$, and in addition, a *screened* Coulomb potential which we can model, *e. g.*, by a repulsive $\varepsilon_l, \varepsilon_{l+1} = V > 0$ possibly extending to the neighbouring sites $l+2$, $l-1$, etc. with an exponential tail. According to the oxygen deficiency in $REBa_2Cu_3O_{7-\delta}$, we can then assume a certain concentration $\delta = n_d/N$ of such defects ($n_d$ the total number of defects), and distribute them randomly.

In Figs. 2, 3, we show the spectra for a chain of length $N = 100$ with $t'/t = 0.1$, and different values of $V$, and $n_d$. A length of $N = 100$ is sufficient, since we are not interested in the very low frequency behaviour. We compared the results for chain lengths of 100, 200, and 400, and found no significant difference at frequencies $\omega \gtrsim t/50$. The precise value of $t'/t$ does not affect the shape of the spectra as long as $t'/t \ll 1$ (which is the physically relevant region). In these calculations, we used a potential which extends to the three sites neighbouring a vacancy falling off exponentially with a screening length equal to the lattice constant.

We always used a band filling of $n := n_e/N = 0.5$, in accordance with the experimentally estimated value mentioned in the introduction. Each curve represents the average over 1000 samples of the disorder, and for clarity of presentation, we introduced a smoothening of the delta functions by averaging the points over small frequency intervals typically of size $t/200$.

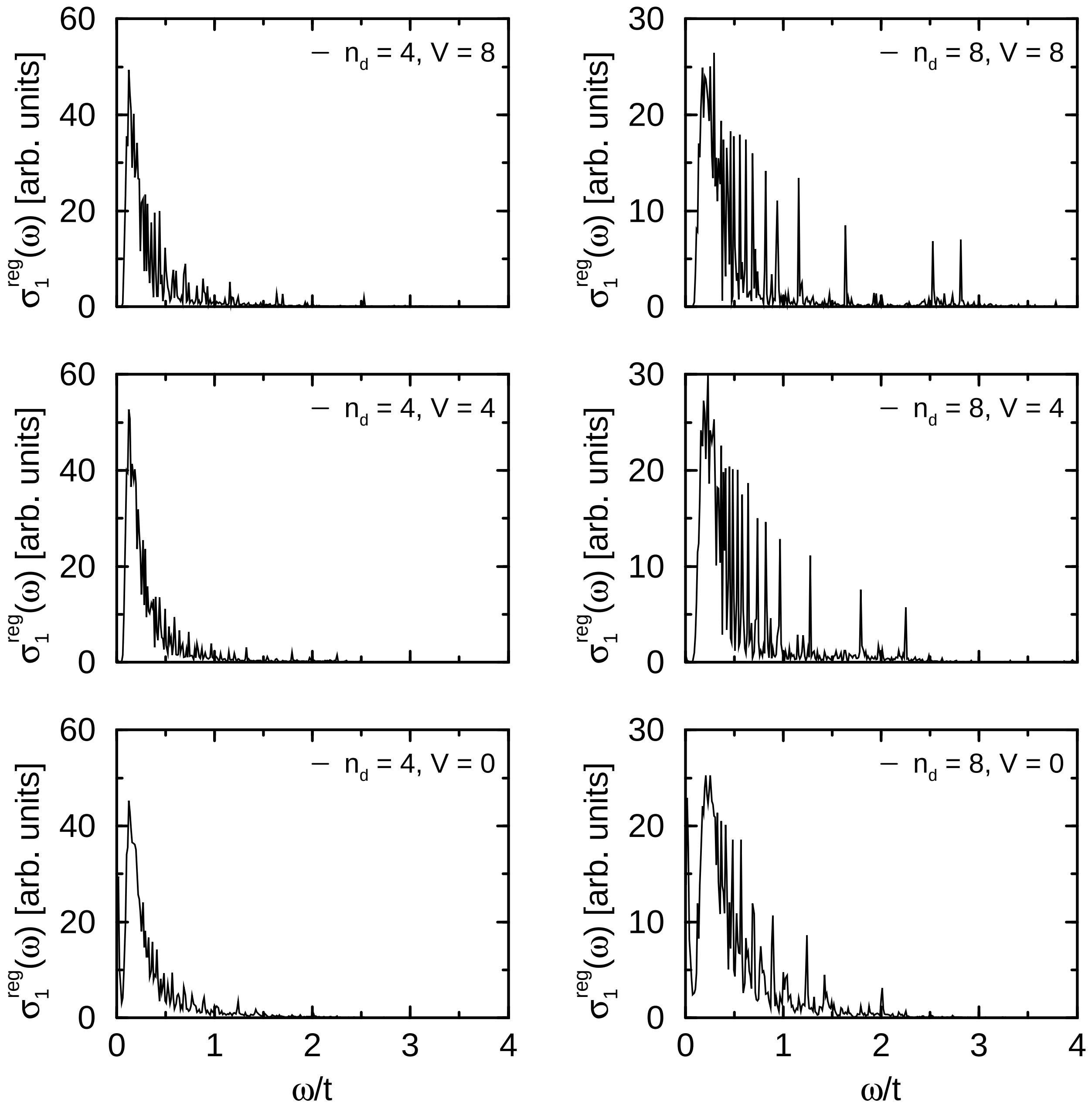

**Figure** 2: Optical conductivity from the model for vacancy disorder. The results were obtained for a chain of length $N$ = 100 at $n$ = 1/2 using the parameter $t'/t$ = 0.1, and different values for $V$, and $n_d$.

The characteristic feature of all spectra is a broad peak at low energy, and a rather rapid fall-off towards higher frequencies. Comparing the results for different values of $V$ with fixed $n_d, t'/t$, we note that the influence of $V$ is not very big, when varied between $V/t = 0 - 8$. The relevant effect seems to result from the reduced $t'$ which introduces strong back-scattering at the vacancies, and therefore effectively cuts the chain into isolated pieces.

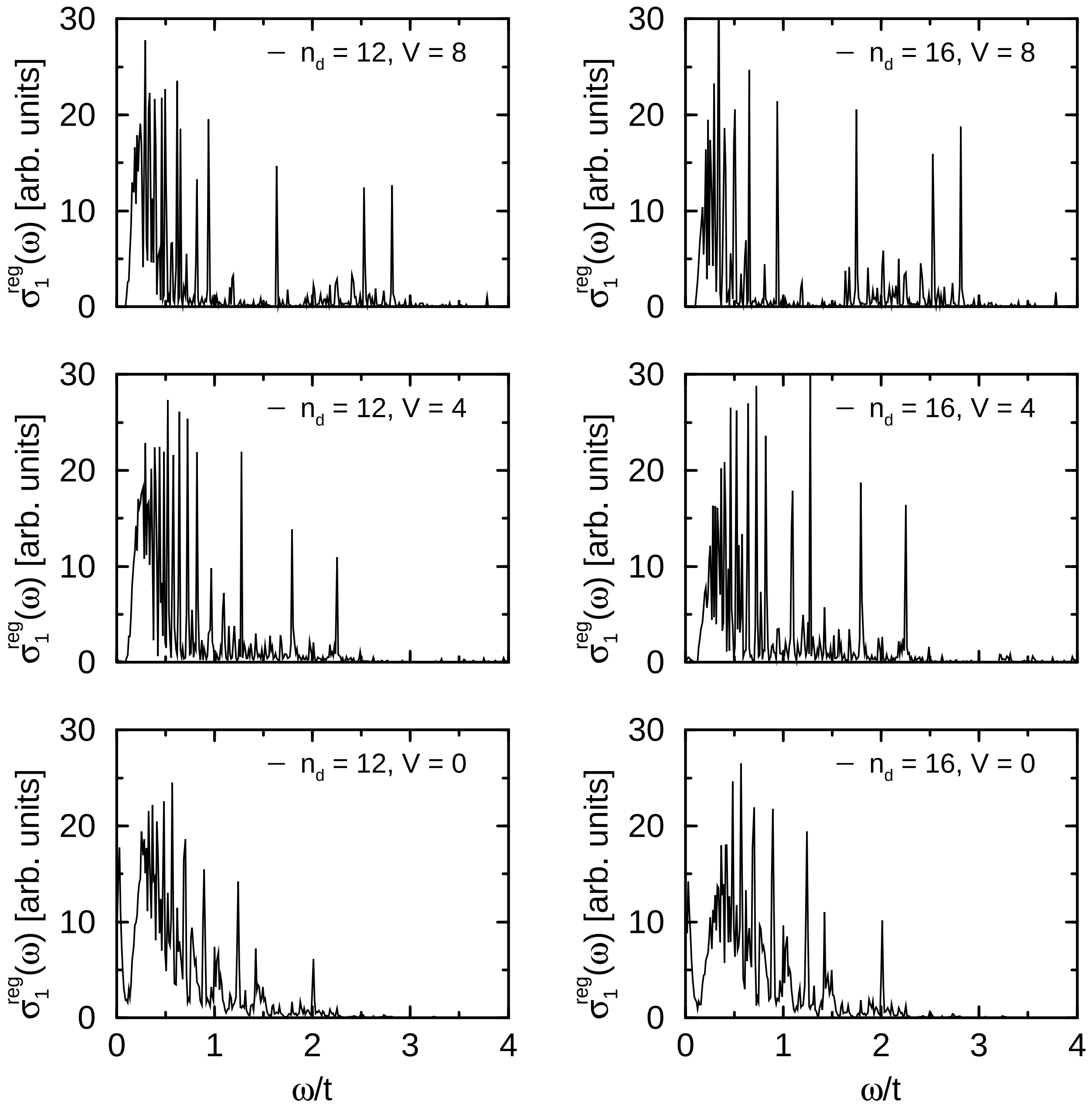

**Figure** 3: Optical conductivity from the model for vacancy disorder. The results were obtained for a chain of length $N = 100$ at $n = 1/2$ using the parameter $t'/t = 0.1$, and different values for $V$, and $n_d$.

As we increase the number of defects $n_d$, we observe that the low-energy peak broadens, and shifts to higher frequencies. This can be explained from the fact that the average length $\overline{N}$ of perfect chain segments decreases as $n_d$ grows, and consequently, the quasi Drude peak at $\omega \approx t/\overline{N}$ of such segments shifts to higher energy. At the same time, the spectra become more and more noisy, dominated by strong isolated peaks corresponding to certain values of $\overline{N}$.

Setting $t = 0.4$eV, we can try a comparison with the experiment. The present model predicts a shift of the observed mid-infrared peak to higher frequencies, when the oxygen concentration is reduced. Since the exact oxygen content (and therefore $n_d$) is difficult to estimate experimentally, we cannot say with certainty whether the absolute value of the calculated peak position is at the correct frequency for a particular sample. However, assuming an upper bound for the oxygen deficiency $\delta < 0.1$ in $REBa_2Cu_3O_7$, the calculated value appears at $\approx 0.1$eV, half of the experimental value. In addition, there is also no sign of a smooth high-frequency tail as observed experimentally. Furthermore, the data in reference [3] taken from samples $YBa_2Cu_3O_{7-\delta}$ with different $\delta$ do not show any strong influence of $\delta$ on the peak position, in contradiction to the calculations. In conclusion, it seems that the oxygen vacancy disorder (at least if modeled as we did) is also not enough to explain the observed mid-infrared peak.

The second kind of disorder that could be present is simple potential disorder, which might originate, *e. g.*, in randomly positioned oxygen interstitials (see Fig. 1). We model this by random on-site energies $\varepsilon_l$ on each site, uniformly distributed in some interval $[-W/2, W/2]$, the size of which is a measure of the disorder strength. In addition, we can also allow for variations in the hopping matrix elements as before. This type of model is often used in the context of Anderson localization, and it is known from various studies [17], that $\sigma(\omega)$ shows a peak at low energy, and a high-frequency Drude-like tail. Nevertheless, we have calculated $\sigma(\omega)$ also for such disorder, since we were interested in the influence of the random $t_l$.

We first study the effect of pure potential disorder, setting $n_d = 0$. Fig. 4 shows the spectra calculated for $N = 100, n_e = 50$, and various values of $W/t$. As expected, we find a peak at low energy, and a subsequent slowly falling high-frequency tail. The position of the peak rapidly shifts to higher frequencies as the disorder strength $W$ is increased. Setting $t = 0.4$eV, we find that a value of $W \approx 1.6$eV is needed, to obtain the experimental peak position of $\omega_{\max} \approx 0.2$eV. In Fig. 5, we superpose the spectrum from $PrBa_2Cu_3O_{7-\delta}$ with the calculated one at $W/t = 4$. The agreement in both, the shape and position of the peak, as well as the high-frequency tail is astonishing. From a purely theoretical point of view, it therefore seems that this model is fully adequate to describe the optical conductivity of the $CuO_3$-chains. However, doubts concerning the validity of this scenario arise because of the large disorder strength $W \approx 1.6$eV required to fit the experiment. We shall comment on this point at the end of this section.

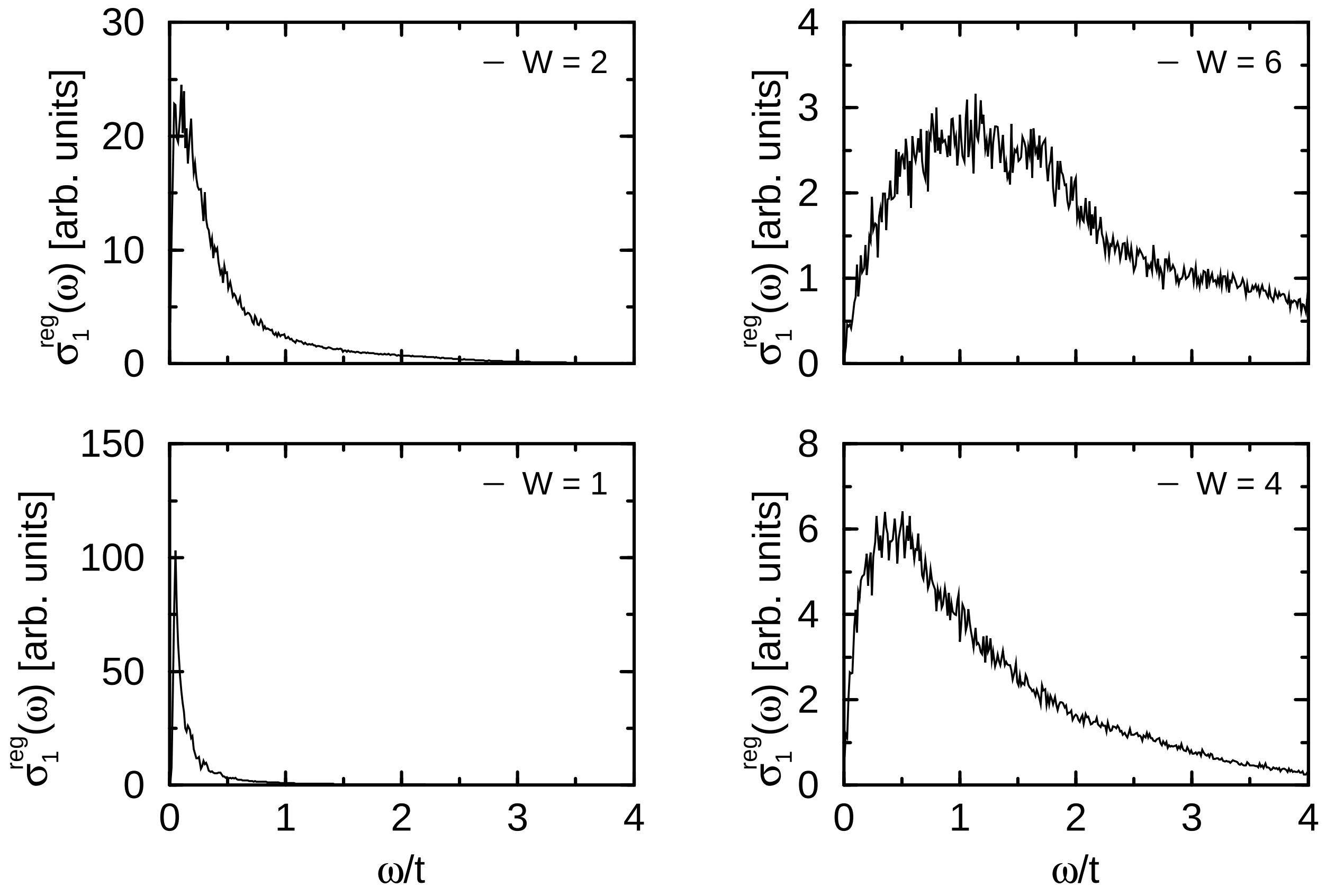

**Figure** 4: Optical conductivity from the model for pure potential disorder. The results were obtained for a chain of length $N = 100$ at $n = 1/2$, and different values for the disorder strength $W$.

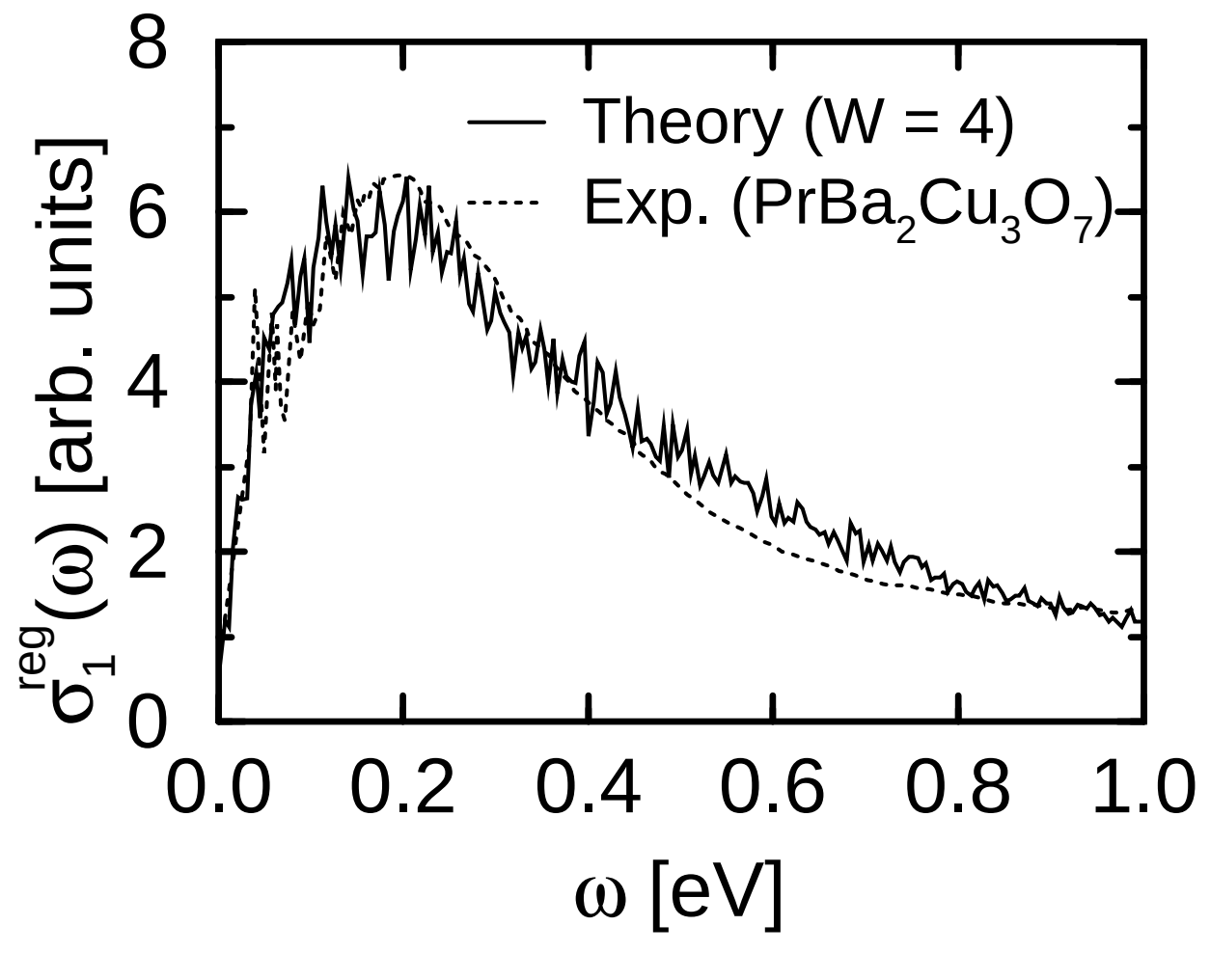

**Figure** 5: A comparison of the experimental spectrum obtained in Ref. [1] with the curve calculated using the model with pure potential disorder, and disorder strength $W/t = 4$.

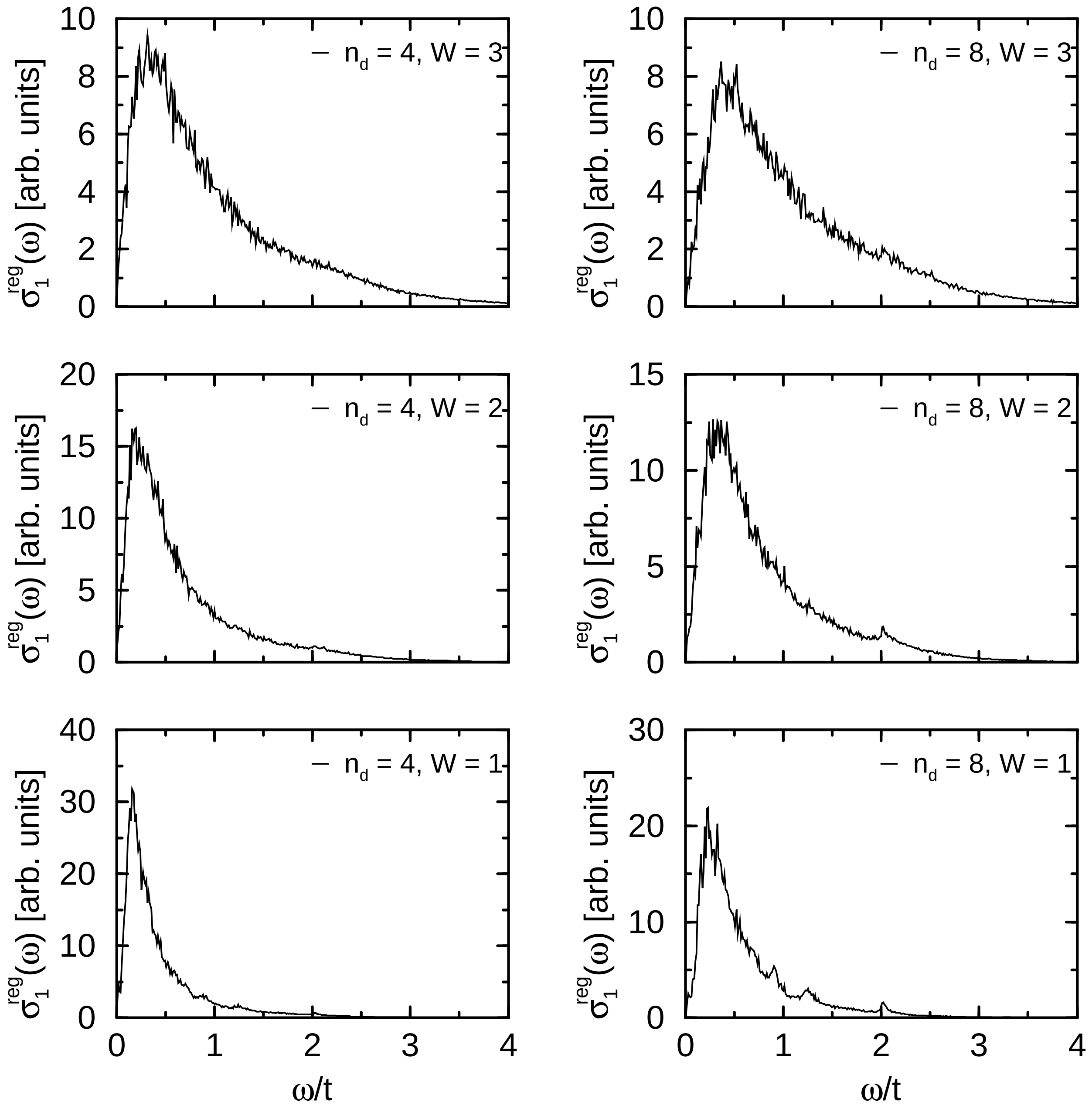

**Figure** 6: Optical conductivity from the model with potential and vacancy disorder. The results were obtained for a chain of length $N = 100$ at $n = 1/2$, using different values for the disorder strength $W$, and vacancy number $n_d$.

Let us now proceed by including vacancy disorder, in this case restricting ourselves to values of $t'/t = 0.1, V = 0$, since the effect of a finite $V$ was found to be not so crucial. Figs.

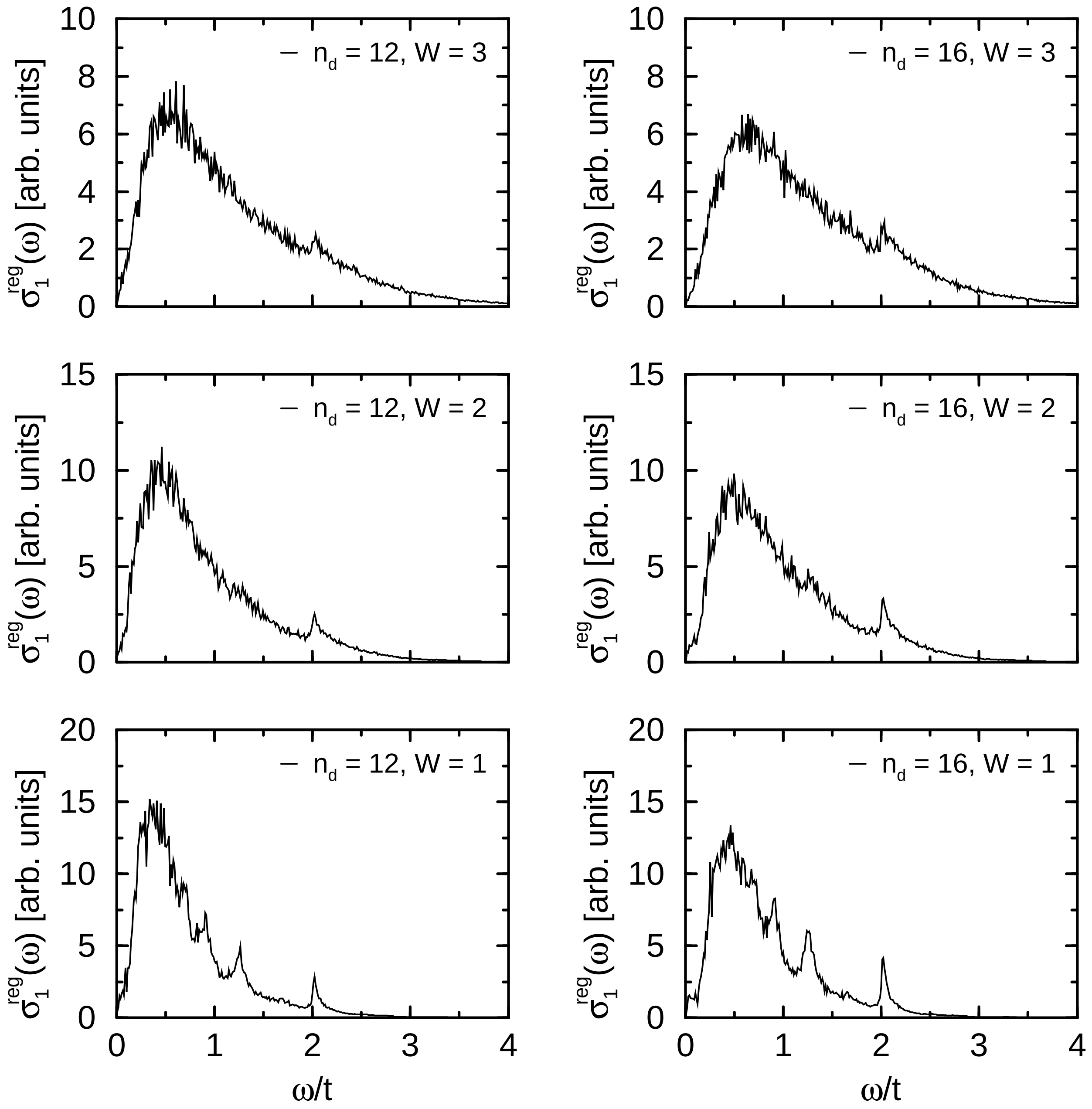

**Figure** 7: Optical conductivity from the model with potential and vacancy disorder. The results were obtained for a chain of length $N = 100$ at $n = 1/2$, using different values for the disorder strength $W$, and vacancy number $n_d$.

6, 7, show the spectra for defect numbers $n_d = 4, 8, 12, 16$, and different strengths of potential disorder $W$. If we compare different values of $n_d$ for fixed $W$, we observe that the peak shifts to larger energies as $n_d$ is increased (as expected). At the same time, for smaller values of $W/t \lesssim 2$, and $n_d \gtrsim 8$, we recover in the high frequency tail, the structure of the isolated quasi Drude peaks which correspond to very short chain segments.

A comparison of *e. g.*, the $n_d = 8$ calculation to the Pr spectrum, shows that now only a value of $W \approx 1.2$eV is needed to obtain a fit as good as the one for $n_d = 0, W/t = 4$. However, going to an even larger $n_d = 16$, the correct peak position requires a value of $W/t \approx 1$, in which case the high-frequency tail is not very well fitted anymore, especially regarding the strong quasi Drude peaks in the calculated spectrum.

In summary, a model including both potential and vacancy disorder exhibits an optical conductivity spectrum which is extremely close to the measured one. However, the potential disorder strength needed to fit the experiment, $W \approx 1.2 - 1.6$eV, is rather large, and the source of such a strong random potential on *each* site is not obvious. A possible type of defect producing such a strong potential could be an oxygen interstitial between two Cu(1) belonging to *neighbouring* chains (see Fig. 1). The corresponding electrostatic potential energy $V_c = e^2/(\varepsilon\, r)$ can be estimated to be $V_c \approx 1.2$eV (using a rather small value for the dielectric constant $\varepsilon = 5$) at the regular oxygen sites nearest neighbour to the interstitial. Nevertheless, it seems a rather high density of such defects would be necessary to produce the required random potential on *each* site.

In the next chapter, we shall investigate a third mechanism which could give rise to the observed optical response, namely polaronic electron-phonon coupling.

## IV. Polaronic Electron-Phonon Coupling

It is well known from various studies of small polaron physics, that systems with a strongly localized electron-phonon (e-ph) interaction usually show a infrared absorption band. Therefore it is worthwhile to look more closely at possible phonon modes in the 1-2-3 materials, which might give rise to such a coupling. The standard theoretical approaches to these systems, employ some kind of mean-field decoupling to handle the polaron-polaron interactions. This can give satisfactory results in the limit of small carrier density, or coupling. However, in the case of interest, the carrier density $n \approx 0.5$/Cu is not low, and even close to a commensurate value at which it can be favourable to form some kind of a lattice superstructure, *i. e.,* a charge density wave (CDW). Furthermore, as we shall show below, the estimated strength of the e-ph coupling in the $CuO_3$-chains puts us in a regime of intermediate coupling, which is difficult to handle with standard perturbative methods. Hence, we need a technique, which is able to take into account the effect of such correlations in a satisfactory way.

We start from the 1D $t$-$J$-model as the effective electronic model having been motivated above. Bearing in mind that an empty site in the $t$-$J$-model corresponds to a $Cu^{III}$ singlet formed from Cu $3d_{x^2-y^2}$ and O $2p_{y,z}$ holes, we can identify an effective way of coupling the apical oxygen O(4) vibrational mode with displacement in the **c**-direction to the holes in the singlet (see Fig. 8): The energy of the singlet is mainly governed by the two hopping matrix elements $t_{pd}$ between Cu $3d_{x^2-y^2}$ and O $2p_{y,z}$ orbitals, and $t_{pp}$ between neighbouring O $2p_{y,z}$ orbitals. As the two apical O atoms above and below a Cu are moved towards it, the values of $t_{pd}, t_{pp}$ increase, and the total energy of the singlet is lowered due to the gain in kinetic energy. Taking for the singlet the approximate wave function

$$|\mathrm{Cu^{III}}\rangle = \left[a_0\, c^\dagger_{d\uparrow}c^\dagger_{d\downarrow} + a_1\left(c^\dagger_{d\uparrow}c^\dagger_{p\downarrow} - c^\dagger_{d\downarrow}c^\dagger_{p\uparrow}\right) + a_2 c^\dagger_{p\downarrow}c^\dagger_{p\uparrow}\right]|\rangle, \tag{18}$$

which is localized on a single Cu and its four surrounding oxygens, we can calculate the total energy of the singlet $E_{\mathrm{singlet}}$ as a function of the displacement $Q$ (measured from the experimental Cu(1)-O(4) distance) using the empirical scaling laws for the hopping matrix elements [18]

$$\frac{t_{pd}(r_{pd})}{t_{pd}(r'_{pd})} = \left(\frac{r'_{pd}}{r_{pd}}\right)^{7/2} \quad ; \quad \frac{t_{pp}(r_{pp})}{t_{pp}(r'_{pp})} = \left(\frac{r'_{pp}}{r_{pp}}\right)^{2}, \tag{19}$$

where $r_{pd}, r_{pp}$ are the relevant interatomic distances between atoms with $d$ or $p$ orbitals.

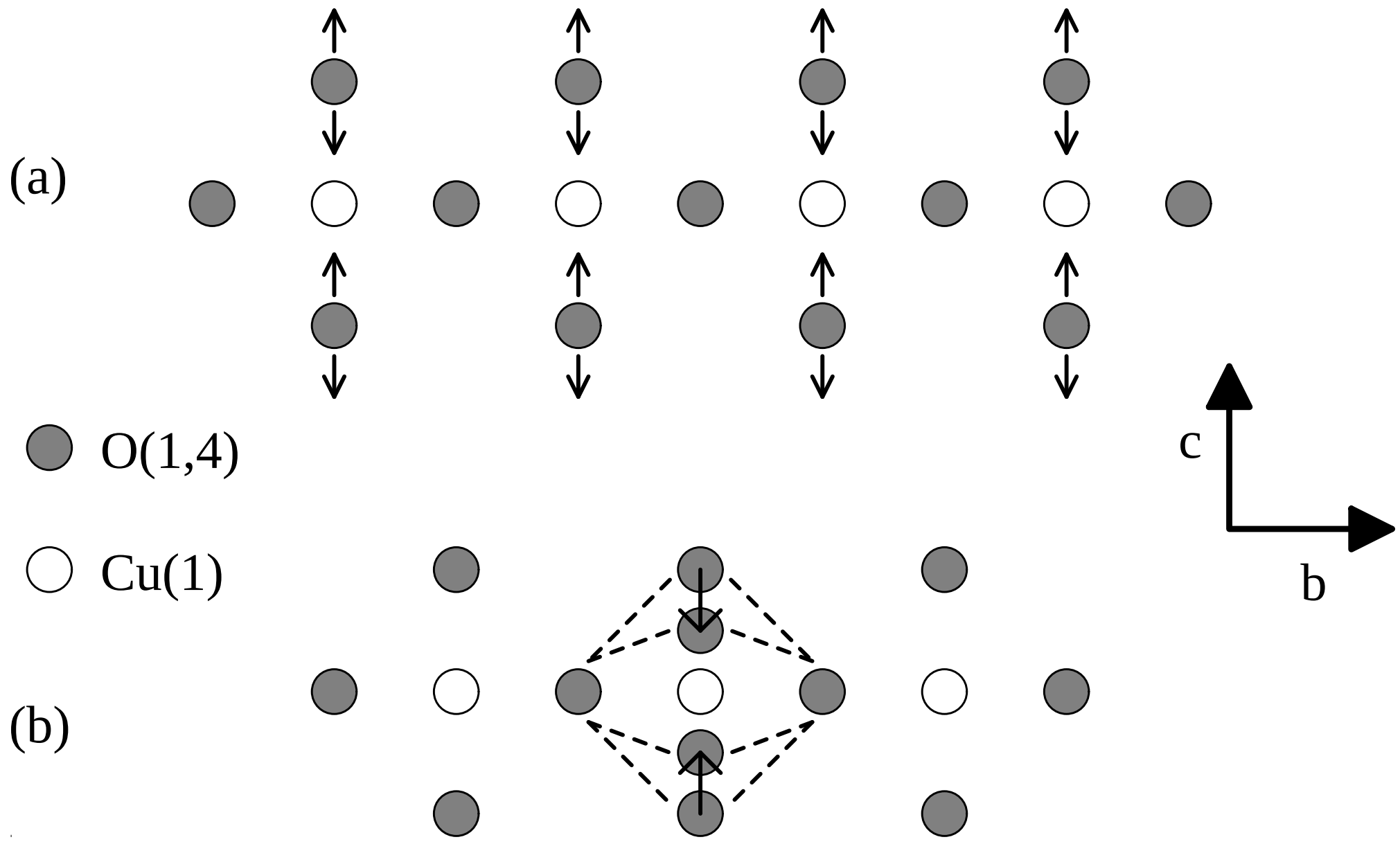

**Figure** 8: Schematic diagram of (a) the vibrational mode which involves displacement of the apical oxygen in the **c**-direction towards the neighbouring Cu(1)-site, and (b) the coupling of this mode to the $Cu^{III}$ singlet.

From the plot of $\Delta_{\text{singlet}}(Q) := E_{\text{singlet}}(Q) - E_{\text{singlet}}(Q = 0)$ in Fig. 9, we note that for values $Q \leq 0.1$Å, the curve is pretty linear, and only for larger $Q$, the curvature becomes important. In view of this, it seems appropriate to model the e-ph interaction by an additional Holstein-like term in the Hamiltonian, which couples the local displacement $Q_l$ linearly to the local hole density $1 - n_{ls}$

$$H_{\text{e-ph}} = \lambda \sum_{ls} Q_l (1 - n_{ls}). \tag{20}$$

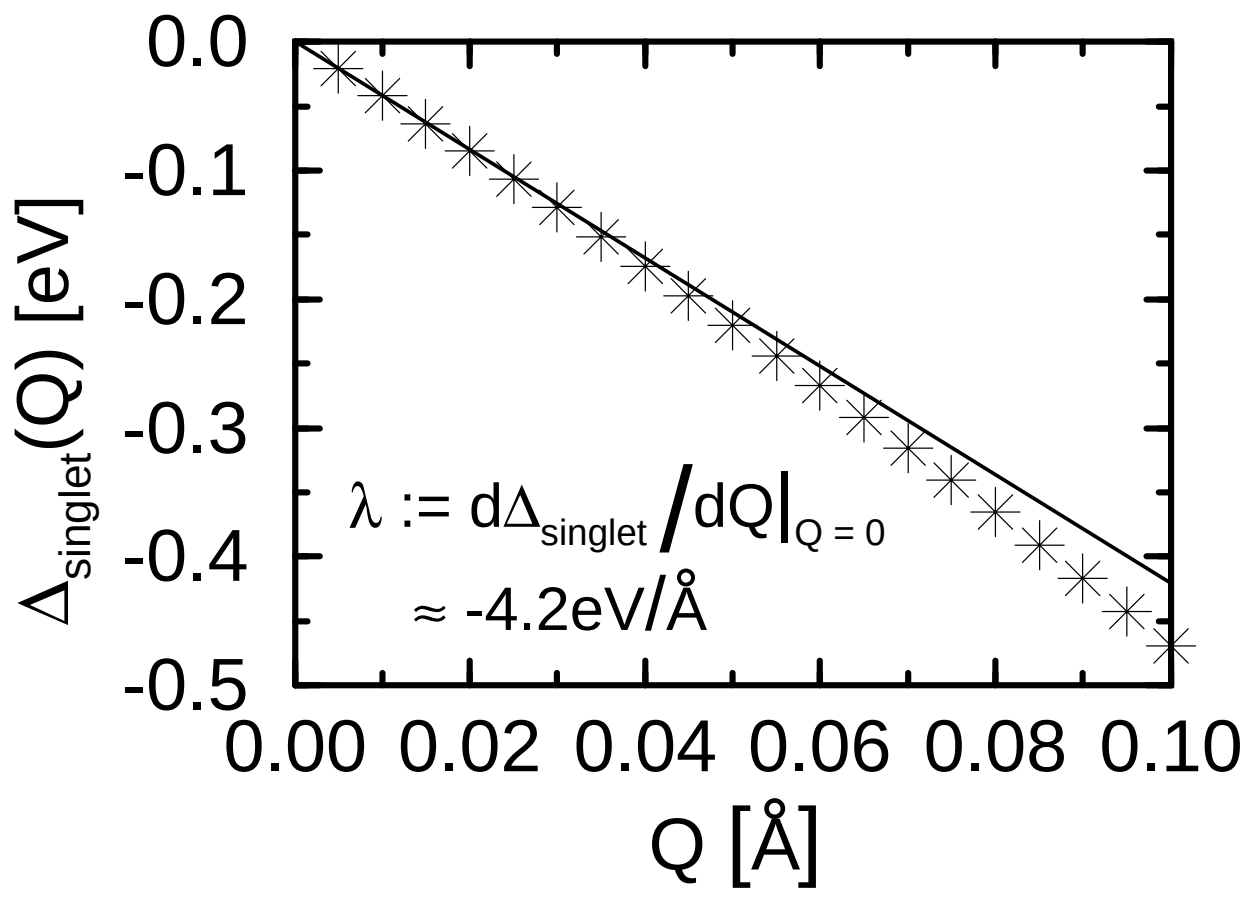

**Figure** 9: The change $\Delta_{\text{singlet}}$ in total energy of the $Cu^{III}$ singlet state as a function of the displacement $Q$ of the apical oxygen as shown in Fig. 8. The gain in kinetic energy of the singlet leads to an estimate of the e-ph coupling constant of $\lambda \approx -4.2$eV/Å.

We have chosen the coupling to the hole, because in the *t-J* model, the unoccupied site corresponds to the $\mathrm{Cu^{III}}$ singlet for which the coupling should be much stronger than for the $\mathrm{Cu^{II}}$ configuration which corresponds to an occupied site. The contribution to the coupling constant $\lambda$ from this mechanism can be obtained from the slope of the curve in Fig. 9 as $\lambda := d\Delta_{\text{singlet}}(Q)/dQ|_{Q=0} \approx -4.2\text{eV/Å}$, where we defined $Q$ to be positive for displacements of the O(4) towards the Cu(1) atoms. However, this value does not contain the purely electrostatic interaction energy, which should have opposite sign. We can estimate it to be of the order of $1-2\text{eV/Å}$. A rough value for the total $\lambda$ can thus be given as $\lambda \approx 2-3\text{eV/Å}$.

The phonon part of the Hamiltonian is the usual one (neglecting dispersion)

$$H_{\text{ph}} = \sum_l \left( \frac{P_l^2}{2M} + \frac{KQ_l^2}{2} \right) \tag{21}$$

with the reduced oxygen mass $M = M_{\text{Ox.}}/2 \simeq 8\text{amu}$, and the spring constant $K$. In terms of the phonon creation and annihilation operators $a_l^\dagger, a_l$

$$Q_l = \left( \frac{\hbar}{2M\Omega} \right)^{1/2} (a_l + a_l^\dagger), \qquad P_l = i \left( \frac{\hbar\Omega M}{2} \right)^{1/2} (a_l - a_l^\dagger), \tag{22}$$

with the phonon frequency $\Omega = \sqrt{K/M}$, the total Hamiltonian becomes

$$H = H_{t-J} + \hbar\Omega \sum_l a_l^\dagger a_l + \Lambda \sum_{ls} (1 - n_{ls})(a_l + a_l^\dagger), \tag{23}$$

and the new coupling constant is $\Lambda = (\hbar\lambda^2/2M\Omega)^{1/2}$. The phonon frequency for the mode under consideration is known rather accurately from Raman scattering [19, 20] as $\hbar\Omega \approx 60-65\text{meV}$, depending on the material. This allows for an estimate of the spring constant as $K = \Omega^2 M \approx 7\text{–}8\text{eV/Å}^2$. The polaron binding energy $E_B := \lambda^2/(2K) = \Lambda^2/(\hbar\Omega)$, which is the most convenient way to measure the effective interaction strength, is therefore of the order of $E_B \approx 0.25\text{–}0.64\text{eV}$. This is a value comparable to the effective bandwidth $4t \approx 1.6\text{eV}$, and therefore puts us in a regime of *intermediate coupling*, which is the most difficult to treat analytically. Note, that our choice of coupling the phonon displacement to the local hole rather than electron density does not affect any properties of the e-ph Hamiltonian (23). Apart from a constant energy shift, the Hamiltonian is electron-hole symmetric (as long as the kinetic energy part is).

The spinless version of the Hamiltonian (23) is the usual Holstein or molecular-crystal model [21]. The optical conductivity of this model has been studied in the weak [22, 23] as well as in the strong coupling case [24] using perturbative methods. However, in none of the approaches used were the polaron-polaron interactions taken into account, *i. e.*, the results are only trustworthy (especially in the strong coupling limit) in the single polaron limit.

There are a number of reasons, why we consider these approaches do not to apply for our problem: (i) We are interested in a situation with high carrier concentrations, *i. e.*, polaron-polaron interactions are important, particularly in the vicinity close to density $n = 1/2$. (ii) The estimated value of the e-ph coupling corresponds neither to weak nor strong coupling, so there is no guarantee that perturbation series rapidly converge. (iii) We are also interested in the effect of the spin degrees of freedom, even though we do not expect them to be very important at realistic values of the spin-exchange constant. (iv) We would like to include the effect of vacancy disorder by choosing open boundary conditions as already explained in section III.

An analytical treatment in the interesting parameter regime including the additional effects of spin-exchange coupling *and* disorder is hopelessly difficult. Consequently, we chose to study the model using a numerical technique, the Lanczos method. In an accompanying paper [25], we present a detailed discussion of our numerical results for the simple Holstein model and contrast them with the analytic results mentioned above.

# V. Numerical treatment of the *t*-*J* Holstein Hamiltonian

A numerical method suitable for dealing with strong correlations would in principle be the quantum Monte Carlo (QMC) method. This technique was used by Hirsch and Fradkin [26] to investigate the CDW transition at half filling. They presented strong evidence that for arbitrary finite values of the phonon frequency $\Omega$, the spinless Holstein model exhibits a transition from a metallic to a CDW state at a finite critical value of the e-ph coupling constant $E_B^c$. They derived the $\tilde{t}$-$\tilde{V}$ model as the effective second order Hamiltonian in the strong coupling limit, and showed that the ratio of the effective parameters $\lim_{E_B \to \infty} \tilde{t}/\tilde{V} = 0$. We recall, that most standard analytical approaches only take into account the hopping renormalization from the first order correction, which defines a small polaron band with a bandwidth $\sim \tilde{t} = te^{-E_B/\Omega}$. As Hirsch and Fradkin noted, the fact that the $\tilde{t}$-$\tilde{V}$ model exhibits a CDW transition at $\tilde{t}/\tilde{V} = 1/2$, gives strong evidence for an analogous CDW ground state in the strong coupling limit of the Holstein model.

The drawback of the QMC method is that it is not straightforward to calculate dynamical quantities such as $\sigma(\omega)$, because the simulation gives direct results for dynamical quantities only as a function of imaginary frequencies. Therefore one needs to analytically continue to the real axis, which is not always a well-defined procedure. In contrast, the Lanczos method has been successfully applied to calculate $\sigma(\omega)$ at zero temperature for purely electronic Hamiltonians on finite lattices [27]. This algorithm allows for the *exact* determination of the low-lying eigenvalues and -vectors of the Hamiltonian matrix. From the calculated ground state, it is then possible to extract dynamical quantities such as, *e. g.*, the spectral function $A(k, \omega)$, or the optical conductivity $\sigma(\omega)$, in principle with arbitrary accuracy. The big advantage of the method is that the results are exact, the disadvantage is the limitation (imposed by the computer memory available) to rather small lattices, so that often an accurate finite-size scaling is impossible.

The algorithm works as follows: One generates an arbitrary initial vector $|\Phi_0\rangle$, with the only restriction that $|\Phi_0\rangle$ should have a finite overlap with the ground state in the symmetry sector of the Hilbert space one is interested in. Starting from $|\Phi_0\rangle$, one generates an orthonormal set of states $\{|\Phi_n\rangle\}$ by successively operating with $H$ on $|\Phi_n\rangle, n = 0, \cdots, M$, and orthonormalizing to all the previous vectors, such that the Hamiltonian is represented by a real symmetric tridiagonal matrix in the basis $\{|\Phi_n\rangle\}$. In more detail, one obtains

$$H|\Phi_0\rangle = \alpha_0|\Phi_0\rangle + \beta_0|\Phi_1\rangle, \tag{24}$$

$$\alpha_0 = \langle\Phi_0|H|\Phi_0\rangle, \qquad \beta_0 = \sqrt{\langle\Phi_0|H^2|\Phi_0\rangle - \alpha_0^2}.$$

The $n$th iteration is

$$H|\Phi_n\rangle = \beta_{n-1}|\Phi_{n-1}\rangle + \alpha_n|\Phi_n\rangle + \beta_n|\Phi_{n+1}\rangle, \tag{25}$$

$$\alpha_n = \langle\Phi_n|H|\Phi_n\rangle, \qquad \beta_n = \sqrt{\langle\Phi_n|H^2|\Phi_n\rangle - \alpha_n^2 - \beta_{n-1}^2},$$

In practice, these operations are performed on a computer. The main difficulty is to find an effective way for obtaining $H|\Phi_n\rangle$ from $|\Phi_n\rangle$.

The extreme eigenvalues and -vectors of the such generated $M \times M$ tridiagonal matrix converge extremely fast to the real ones of $H$. Starting from a completely random $|\Phi_0\rangle$ (which is the safest choice), one typically needs $M = 60 - 100$ iterations to obtain an accuracy to 12 decimal places in the lowest eigenvalue $E_0$, and an error $\Delta = (\langle\Psi_0|H - E_0|\Psi_0\rangle)^{1/2} \approx 10^{-7} - 10^{-8}$ for the ground state $|\Psi_0\rangle$. The number of iterations needed for convergence of the ground state energy depends strongly on the splitting $E_1 - E_0$ to the first excited state, and increases significantly as the two levels get close.

For the calculation of the optical conductivity at zero temperature, one usually employs the continued fraction expansion of the Greens function [27]

$$G(Z) = \langle\Psi_0|\, j(Z-H)^{-1} j|\Psi_0\rangle = \cfrac{\langle\Psi_0|\, jj\,|\Psi_0\rangle}{Z - \tilde{\alpha}_0 - \cfrac{\tilde{\beta}_1^2}{Z - \tilde{\alpha}_1 - \cfrac{\tilde{\beta}_2^2}{Z - \tilde{\alpha}_2 - \cdots}}}\quad , \tag{26}$$

where the coefficients $(\tilde{\alpha}_n, \tilde{\beta}_n)$ of the continued fraction are the usual Lanczos coefficients obtained from a second Lanczos run using the starting vector $|\tilde{\Phi}_0\rangle = j|\Psi_0\rangle$. The continued fraction is truncated after $M$ iterations (typically $M = 300$), when a good convergence of the spectrum is obtained. The real part of the optical conductivity is then given by

$$\sigma_1^{\rm reg}(\omega) = -\frac{1}{\pi\omega}\mathrm{Im}\; G(\omega + E_0 + i\delta), \tag{27}$$

where we introduced a small imaginary part $i\delta$, which gives a finite Lorentzian width to the delta peaks appearing in the original definition (6).

Unfortunately, the Lanczos method is not applicable to an e-ph system without further approximation. The reason is simply the fact that the phonon Hilbert space is infinite-dimensional already for a single phonon mode, whereas the Lanczos algorithm can only be used for finite-dimensional Hilbert spaces, since convergence to the true ground state is only guaranteed if the vectors $|\Phi_n\rangle = \sum_m a_{nm}|\phi_m\rangle$ are expressed in a *complete* set of basis states $|\phi_m\rangle$ of the full (in each symmetry sector) Hilbert space. This requires computer storage of the coefficients $a_{nm}$ of at least two vectors at one time. For an infinite-dimensional Hilbert space, clearly, one would require an infinite amount of computer memory in order to obtain exact results.

Therefore, the only possible way to apply the Lanczos method to an e-ph Hamiltonian is to use a variational approximation for the phonon Hilbert space, which makes it finite-dimensional. Since variational approximations are most often motivated by physical intuition or/and computational convenience, obviously, there are many possible choices for a variational ansatz. For our Hamiltonian, we considered the three 'most natural' variational approximations to the phonon Hilbert space, which are within the limits set by current computer power and the Lanczos algorithm.

The basic computational limitation we encounter is the maximum number of independent phonon states that can be allowed for each mode. If we denote this number by $Q$, then on an $N$-site lattice, the total number of linearly independent phonon states is given by $R_{\mathrm{ph}}(N, Q) = Q^N$. On the other hand, the size of the electron Hilbert space (neglecting spin) on a $N$-site lattice with $n$ electrons is given by $R_{\mathrm{e}}(N, n) = \binom{N}{n}$, so that the total Hilbert space has the dimension $R_{\mathrm{tot}}(N, n, Q) = Q^N = \frac{N!Q^N}{n!(N-n)!}$. As an illustration, in table I, we show this number for different values of $N, Q$, and fixed $n = N/2$, where the electron Hilbert space is maximal for a given $N$. Bearing in mind the size of core memory of present supercomputers (~ 1GByte, corresponding to a maximum size $R_{\mathrm{tot}}(N, n, Q) \approx 4.5 \cdot 10^7$), we can extract from these numbers that the maximum tractable lattice sizes are $N = 15$ for $Q = 2$, $N = 9$ for $Q = 4$, $N = 7$ for $Q = 6$, and $N = 6$ for $Q = 8$. We have performed calculations mainly for $Q = 2$. The quality of this approximation, *i. e.*, the influence of additional phonon degrees of freedom, was then tested by a comparison with results for $Q = 4$. As we shall see later, the qualitative features obtained from the $Q = 2$ calculation are not altered. The lattice sizes for $Q > 4$ that are tractable are too small to extract meaningful results. Therefore we did not consider them.

Restricting ourselves to the case $Q = 2$, there are three plausible choices for the two phonon states of each mode. A detailed analysis of the results within these three subspaces will be published in a separate paper [25]. There, we use several different criteria to test the quality of each Ansatz: (i) The total energy, (ii) the ability to reproduce the exact parameters of the effective Hamiltonian in the strong coupling limit, and (iii) a comparison with available results from quantum Monte Carlo simulations. The result of this analysis is that in the interesting parameter range from intermediate to strong e-ph coupling, there is a distinct subspace, which gives by far the best results. It is spanned by the direct product of the displaced (coherent state)

$$|\widetilde{0}\rangle_l = e^{-\eta^2/2} e^{\eta a_l^\dagger} |0\rangle_l , \tag{28}$$

and the undisplaced oscillator ground state $|0\rangle_l$ at each lattice site. We refer to this variational subspace as Ansatz II.

**Table II**: The dimension of the total Hilbert space $R_{\text{tot}}(N, n, Q)$ as a function of the lattice size $N$, and the allowed number of independent phonon states in each mode $Q$.

| Q | $N=6$ | $N=8$ | $N=10$ | $N=12$ | $N=14$ | $N=16$ |
|---|---|---|---|---|---|---|
| 2 | $1.280 \cdot 10^{3}$ | $1.792 \cdot 10^{4}$ | $2.580 \cdot 10^{5}$ | $3.785 \cdot 10^{6}$ | $5.623 \cdot 10^{7}$ | $8.434 \cdot 10^{8}$ |
| 4 | $8.192 \cdot 10^{4}$ | $4.588 \cdot 10^{6}$ | $2.642 \cdot 10^{8}$ | $1.550 \cdot 10^{10}$ | $9.213 \cdot 10^{11}$ | $5.528 \cdot 10^{13}$ |
| 6 | $9.331 \cdot 10^{5}$ | $1.176 \cdot 10^{8}$ | $1.524 \cdot 10^{10}$ | $2.011 \cdot 10^{12}$ | $2.689 \cdot 10^{14}$ | $3.631 \cdot 10^{16}$ |
| 8 | $5.243 \cdot 10^{6}$ | $1.174 \cdot 10^{9}$ | $2.706 \cdot 10^{11}$ | $6.350 \cdot 10^{13}$ | $1.509 \cdot 10^{16}$ | $3.623 \cdot 10^{18}$ |

In the calculations presented in this paper, we always used Ansatz II. In the strong coupling limit, $t/E_B \rightarrow 0$, this choice guarantees the correct ground state energy, since the corresponding subspace contains the true degenerate ground state manifold which is obtained for $t = 0$. One has to be a little careful with this choice, since the two mentioned basis states are *non-orthogonal*

$$_l\langle 0|\widetilde{0}\rangle_l = e^{-\eta^2/2} := \nu . \tag{29}$$

For the Lanczos-method, it is simpler to deal with orthonormal basis states, hence we choose the combination

$$|\phi_1\rangle_l = \frac{1}{\sqrt{1-\nu^2}} \Big( |0\rangle_l - \nu |\widetilde{0}\rangle_l \Big) ; \qquad\qquad |\phi_2\rangle_l = |\widetilde{0}\rangle_l . \tag{30}$$

Let us now turn to the calculations of the optical conductivity. As can be seen from Eq. (6), the evaluation of this quantity involves the whole set of excited states of the full Hamiltonian. Consequently, having a good variational approximation of the true ground state (obtained by optimization of the energy within a subspace of the full Hilbert space) does *not guarantee* that the optical conductivity calculated within the same subspace is as accurate. It can not be excluded a priori that excited states $|i\rangle$ *outside* of this subspace have a considerable current matrix element $\langle i|\, j|0\rangle$ with the ground state within this subspace. Obviously, such contributions are lost by our calculation, and it seems necessary to estimate their importance. Within our approach, the leading corrections can be obtained by extending the local phonon basis to additionally include the first excited states $|1\rangle_l, |\widetilde{1}\rangle_l$ of the (un)displaced oscillator. This will be done in the next section.

In the accompanying paper [25], the doping dependence of the optical conductivity is investigated in the presence of PBC. There, we stress the strong influence of the polaron-polaron interactions on $\sigma(\omega)$, which were neglected in previous calculations of this quantity. The generic physical picture that emerges is the following: For coupling strength large enough to result in an ordered CDW-state at half-filling, the conductivity at zero frequency, *i. e.*, the Drude weight, is proportional to the number of electrons (as for a usual metal) at low electron (hole) density $n \ll 1/2$, while at electron densities close to $n = 1/2$ ($|n - 1/2| \ll 1/2$), it is proportional to the number of domain walls, $n_d = 2|n_e - N/2|$ (see Fig. 11).

In the context of a CDW-state, a domain wall can be described as follows: Let us characterize the ordered CDW-state by a periodic function $n(R_l) = 1/2 + \cos(\pi R_l/a)/2$ which mimics the alternating electron occupation on neighbouring sites. In this picture, a domain wall is a defect in the periodic structure, at which the phase $\phi = \pi R_l/a$ suffers a shift of $\pi$. To the right of this defect, all the electrons are shifted by one lattice site (see Fig. 10). If an additional electron (or hole) is introduced into the CDW, it is easy to convince oneself, that this generates precisely two domain walls (at least in the presence of PBC), and they should behave hole-like, *i. e.*, carry positive charge, for $n < 1/2$, and electron-like, *i. e.*, carry negative charge, for $n > 1/2$. Note, that the origin of this behaviour lies entirely in the effective electron-electron interactions mediated by the phonons.

The doping dependence of $\sigma(\omega)$ in this coupling regime also reflects these interactions: At $n = 1/2$, the dc conductivity vanishes, and at finite frequencies $\sigma(\omega)$ shows the characteristic

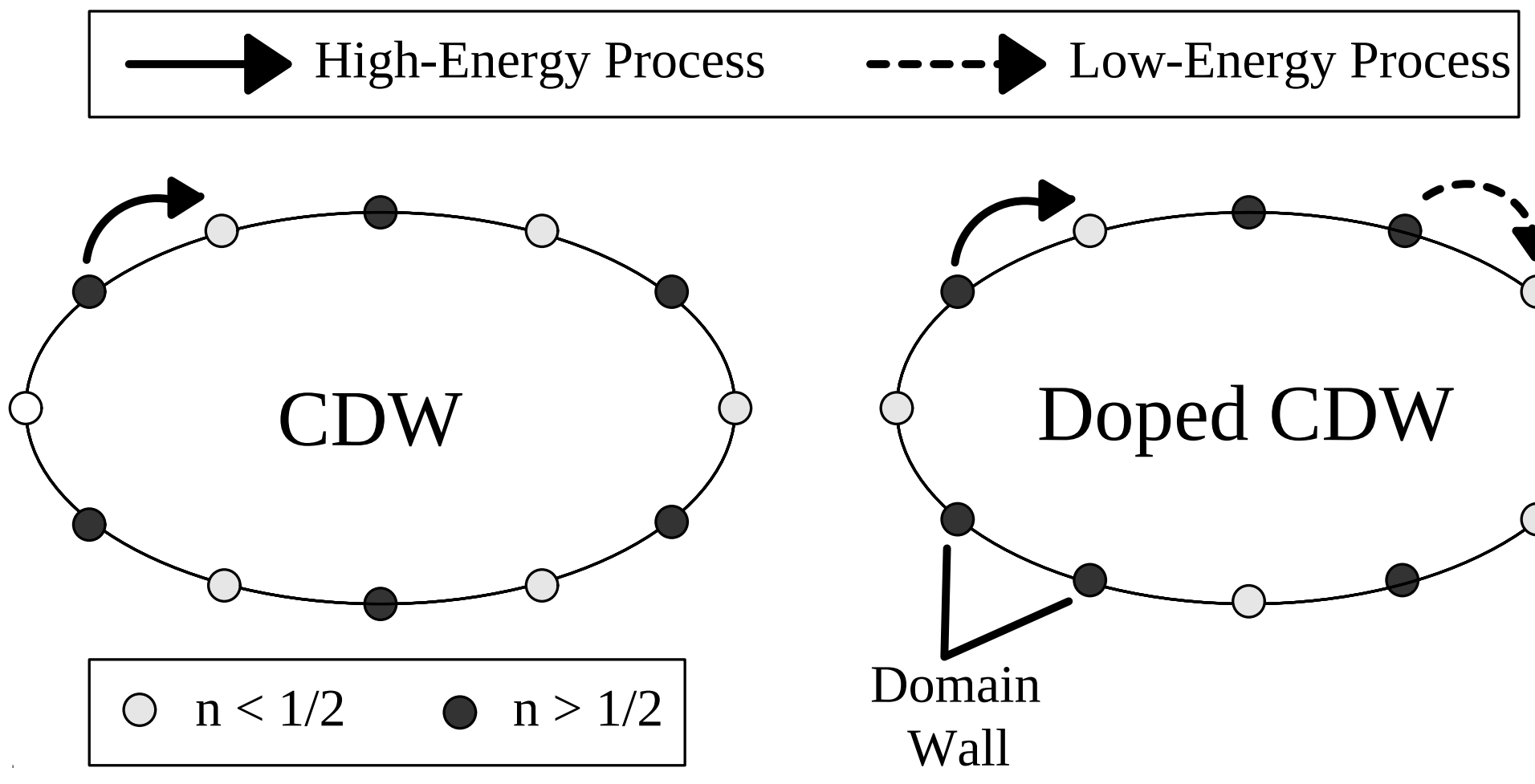

**Figure** 10: Schematic diagram of the dominant excitations induced by an external electromagnetic field in an (un)doped CDW. The doping of the CDW leads to the formation of mobile domain walls.

CDW-gap, $\Delta_{\mathrm{CDW}}$, beyond which a slowly falling high-frequency tail appears (if the coupling is not too large). As we dope away from half-filling, the dc conductivity is reestablished, and the gap disappears. Spectral weight is being transferred from high ($\omega > \Delta_{\mathrm{CDW}}$) to low energy, but the slowly falling high-frequency tail remains. As we approach the low-density limit, the effective interactions become less and less important. This is reflected by the disappearence of the high-frequency tail, and the ratio of the Drude weight to the total spectral weight, which comes close to 1. For small coupling (such that there is no CDW at $n = 1/2$), the doping dependence of $\sigma(\omega)$ is not very pronounced, since the polaron-polaron interactions are negligible. We shall therefore concentrate on the interesting case of sufficiently strong coupling.

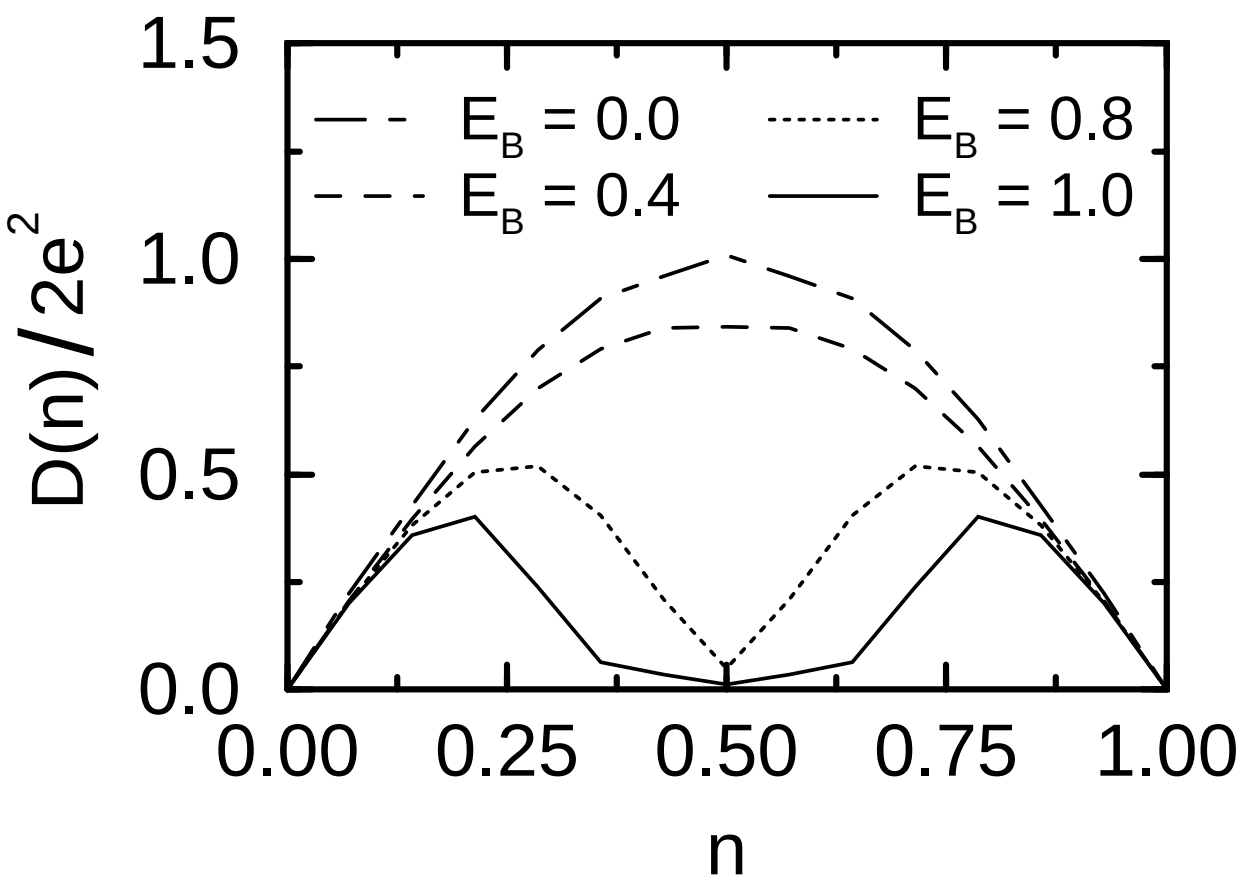

**Figure** 11: Doping dependence of the Drude weight $D$ for various values of the coupling $E_B$ obtained for a lattice with $N = 14$ using (A)PBC at fixed phonon frequency $\Omega = 0.2$

## VI. Comparison with experiment

In this section, we restrict ourselves to a set of parameters for the Holstein Hamiltonian, which represents those expected in the real chains. In section III, we already pointed out the importance of defects in a realistic model for the chains. We also mentioned the experimental finding that the oxygen vacancies on the O(1)-site in the chains, are by far the most important type of defects. Such a vacancy presents a serious perturbation to the electron motion, since it effectively cuts the chains into more or less disconnected segments. In consideration of this fact, we think it is justified to assume that the dominant part of the optical response originates from the excitations in finite chain segments, and that the inter-segment hopping only plays the role of a small perturbation, which we neglect. For our calculations, this means that we should use open boundary conditions (OBC), *i. e.*, an oxygen vacancy is modelled by a *purely reflecting wall*.

We already studied a similar case in section III, where we considered electrons without coupling to phonons. One obvious property of electronic models with OBC is that they cannot support any dc current. Hence, as already mentioned earlier, the Drude weight is transferred to finite frequencies, and especially for rather short chain lengths, this leads to a dramatic change in the low-energy response.

In this section, we shall study the optical conductivity of the Holstein Hamiltonian in the presence of OBC. We shall restrict the parameters to be $t = 0.4$eV, and $\Omega/t = 0.15 - 0.2$, and vary the e-ph coupling strength $E_B$, as well as the band-filling. As for PBC, the largest system we can study for all fillings is the 14-site lattice, so we shall concentrate on this one at the beginning.

In Figs. 12, we show the spectra for electron numbers $n_e = 2$–$7$, phonon frequency $\Omega = 0.2$, and slightly different coupling strengths around $E_B = 0.6 - 0.75$, *i. e.*, values which are very close to the critical one at half filling, $E_B^c(\Omega = 0.2) \approx 0.52$. The common feature of all spectra is a dominant peak, which is slightly shifted away to higher $\omega$ from the frequency (indicated by the dashed vertical lines) at which the non-interacting system shows its main response (see section III). At smaller frequency, there is also considerable absorption, which partly fills in the quasi-gap below the main peak. At low density ($n_e = 2, 3$), there is a rapid fall-off towards higher frequency, whereas for densities close to 1/2, we recover again the slowly falling high-frequency tail, driven by the polaron-polaron interactions. At half-filling, and for $n_e = 6$, the tail follows smoothly onto the main peak, whereas for $n_e = 4, 5$, the peak looks more like an isolated structure superposed with the slowly falling background absorption.

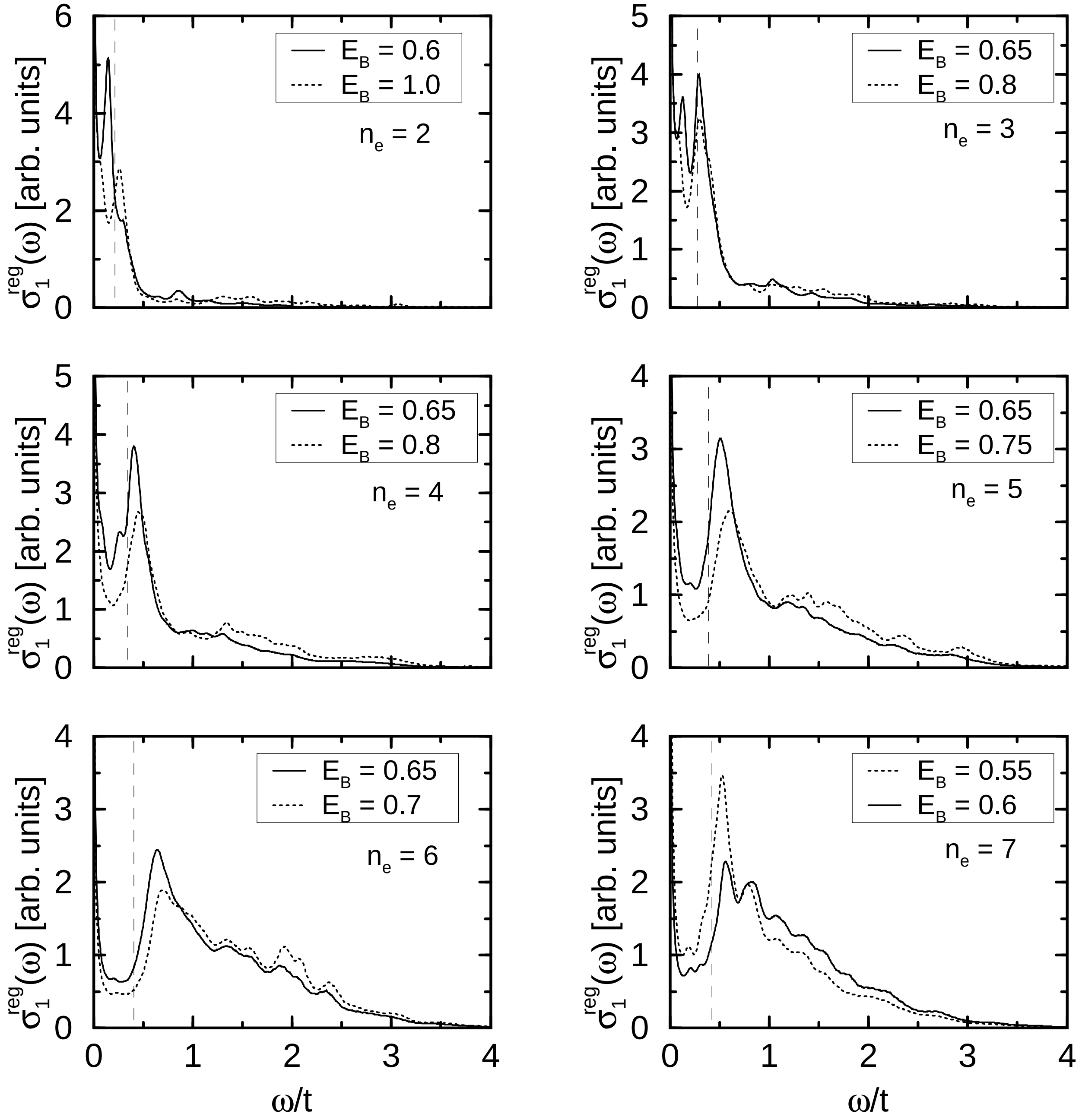

**Figure** 12: The frequency dependence of $\sigma_1^{\mathrm{reg}}(\omega)$ obtained for a lattice with $N = 14$ using OBC. The plots show spectra calculated for different fillings, and coupling strengths $E_B$ (close to $E_B^c$) at fixed phonon frequency $\Omega = 0.2$.

The differences compared to the spectra with PBC in ref. [25] are two-fold: (i) The ap-pearance of the low-energy peak can clearly be related to an optical excitation, which is already present in the non-interacting case, namely the lowest particle-hole excitation of the free electron system. The corresponding energy is largest at half-filling (due to the simple cosine band-structure), and decreases monotonically with the density. This peak dominates the optical conductivity in the presence or absence of the e-ph interactions, at least for not too strong coupling (where it is suppressed for densities close to 1/2). However, in the interacting case, the peak is shifted to higher energies (as compared to the free system), and also broadened as the coupling increases.

(ii) The second new feature is the appearance of spectral weight at very low energy below the dominant peak, which partly fills in the quasi-gap. This weight decreases as a function of increasing coupling in much the same way as the weight under the peak decreases. The combined reduction of weight of these two features can be related to the corresponding decrease of the Drude weight obtained for PBC. Physically, the low-energy weight can be associated with the renormalized charge motion, which occurs on the energy scale $\sim \tilde{t}$. For PBC, this motion contributes to the dc conductivity, whereas here, due to the reflecting walls, the corresponding spectral weight appears at finite frequencies of the order of $\omega \sim \tilde{t}$.

This effect can be observed, when we concentrate on the low-energy spectrum below the main peak, and use a smaller broadening $\delta = 0.01$ for our plots. Fig. 13 shows this part of the spectrum for the case $N = 12$, $n_e = 5$, $\Omega = 0.2$, and three different coupling strengths. Note, how the dominant peak in this low-energy part is shifted towards lower frequencies as the coupling is increased, according to the dependence of $\tilde{t} = te^{-E_B/\Omega}$. A detailed analysis of the peak position revealed exactly the exponential dependence. Furthermore, for strong coupling $E_B > E_B^c$, the inspection of the change in this low-energy spectral weight as a function of doping reveals a very similar dependence as obtained for the Drude weight in the case of PBC (see Fig. 11): For densities close to 1/2, this weight increases with the number of domain walls, whereas for $n \gtrsim 0$, it increases with the number of electrons. In summary, we can say that the low-energy spectrum clearly shows the presence of simultaneous renormalized ($\omega \sim \tilde{t}$), and unrenormalized charge motion ($\omega \sim t$).

In Fig. 14, we show the same spectra for lower phonon frequency $\Omega = 0.15$, and various coupling strengths. The qualitative shape of $\sigma(\omega)$ is very similar to the previous case, but note, that the coupling strength needed to obtain the high-frequency tail is now smaller, because the phonon frequency changes the renormalization factor $\nu^2 = e^{-E_B/\Omega}$.

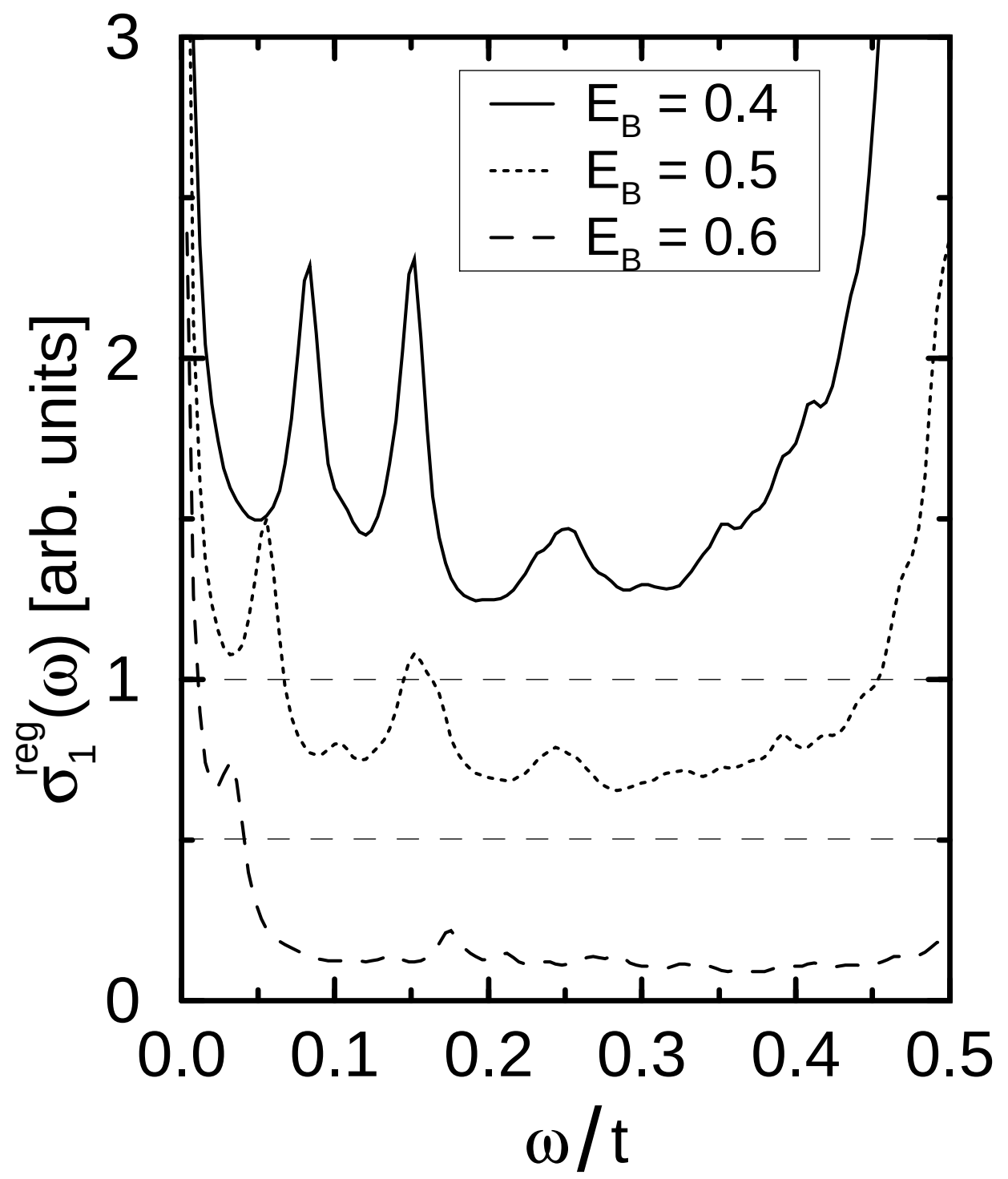

**Figure** 13: The low-frequency part of $\sigma_1^{\mathrm{reg}}(\omega)$ obtained for a lattice with $N = 12$ using OBC at $n_e = 5$. The plots show spectra calculated for different coupling strengths $E_B$ at fixed phonon frequency $\Omega = 0.2$.

The inspection of the spectra at different doping levels showed that the high-frequency tail, which is a characteristic feature of the experimental spectrum, is present only for densities around 1/2. This is a promising feature of the e-ph model, since we know from experimental estimates, that the hole density in the $CuO_3$-chains of both, $PrBa_2Cu_3O_{7-\delta}$ and $YBa_2Cu_3O_{7-\delta}$ is approximately 1/2 per Cu, which translates to the value $n = 1/2$ in our model. Hence, we shall now compare the experimental spectrum of $PrBa_2Cu_3O_{7-\delta}$ with those calculated spectra which give the best agreement.

Fig. 15 shows the experimental $\sigma_1^{\mathrm{reg}}(\omega)$ together with the ones calculated for parameters (a) $N = 14$, $n_e = 6$, $\Omega = 0.2$, $E_B = 0.65$, and (b) $N = 13$, $n_e = 6$, $\Omega = 0.2$, $E_B = 0.55$. We have used the usual value $t = 0.4$eV to fix the energy scale. The agreement of the calculated with the experimental curve is fairly good. In particular, the high-frequency tail is very well approximated by the calculation, and also the ratio $\sigma(\omega = \omega_{\mathrm{peak}})/\sigma(\omega = 1.0\mathrm{eV})$ of the peak height to the weight at the bottom of the tail ($\omega = 1.0$eV) is correct. However, there is some discrepancy at small frequencies: The quasi-gap below the main peak is not observed in the experiment. There, the conductivity drops very sharply when approaching zero frequency.

The total spectral weight can be expressed in terms of an effective carrier number per unit

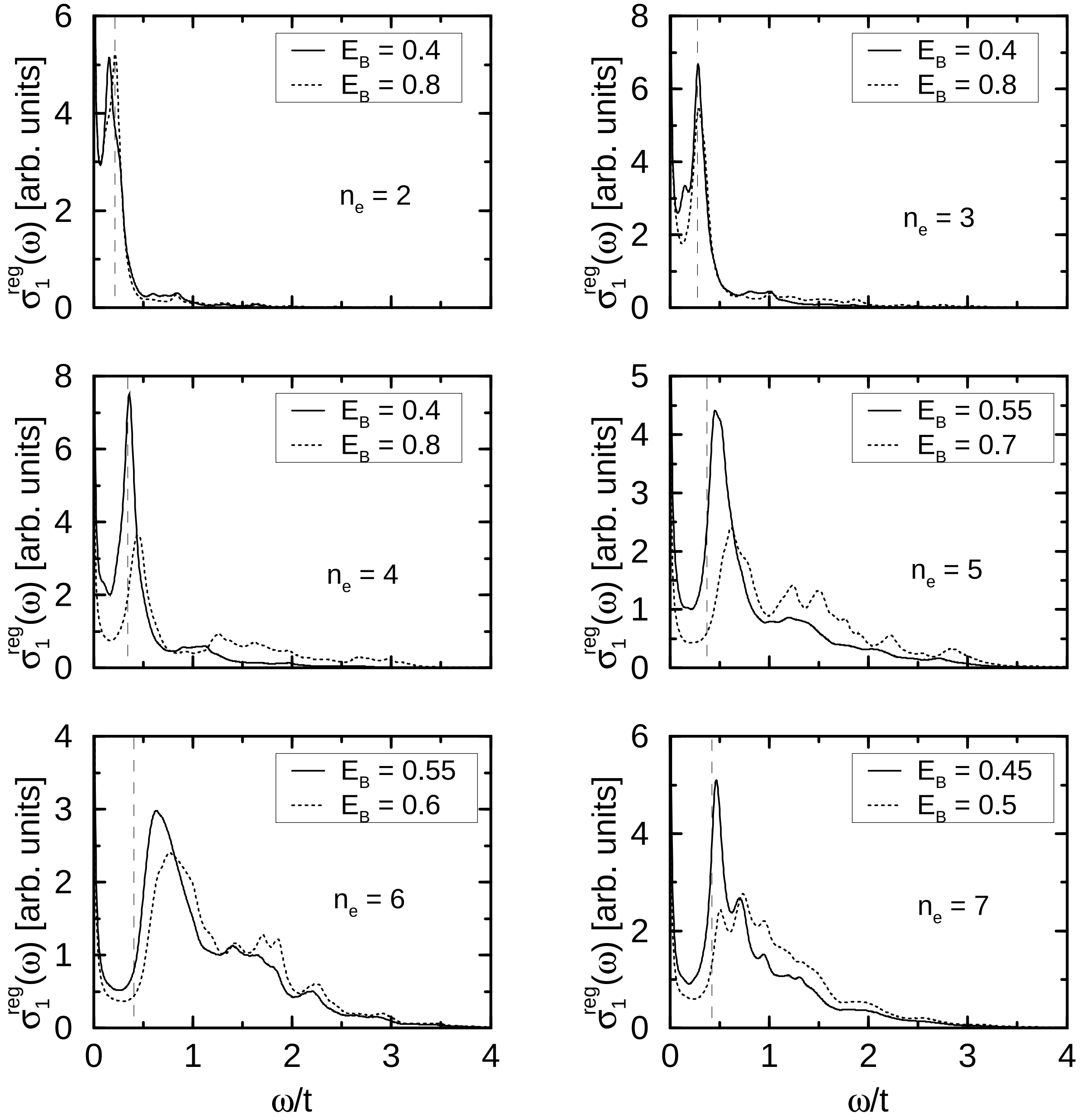

**Figure** 14: The frequency dependence of $\sigma_1^{\text{reg}}(\omega)$ obtained for a lattice with $N = 14$ using OBC. The plots show spectra calculated for different fillings, and coupling strengths $E_B$ (close to $E_B^c$) at fixed phonon frequency $\Omega = 0.15$.

cell

$$n_{\text{eff}} := \frac{2m_0 V_{\text{cell}}}{\pi e^2} \int_0^\infty d\omega\, \sigma(\omega) \quad \left( = \frac{m_0 a_0^2}{\hbar^2} \frac{\langle -T \rangle}{N} \text{ for tight-binding models} \right), \tag{31}$$

where $m_0$ the free electron mass, $V_{\text{cell}}$ the volume of the unit cell, $e$ the elementary charge, and $a_0$ the lattice constant along the chain, *i. e.*, the **b**-direction. From our calculations Fig. 15, we obtain a value of $n_{\text{eff}} \approx 0.43 - 0.45$. This compares rather well with the experimental one $n_{\text{eff}} \approx 0.5$. The discrepancy comes from the low-frequency part.

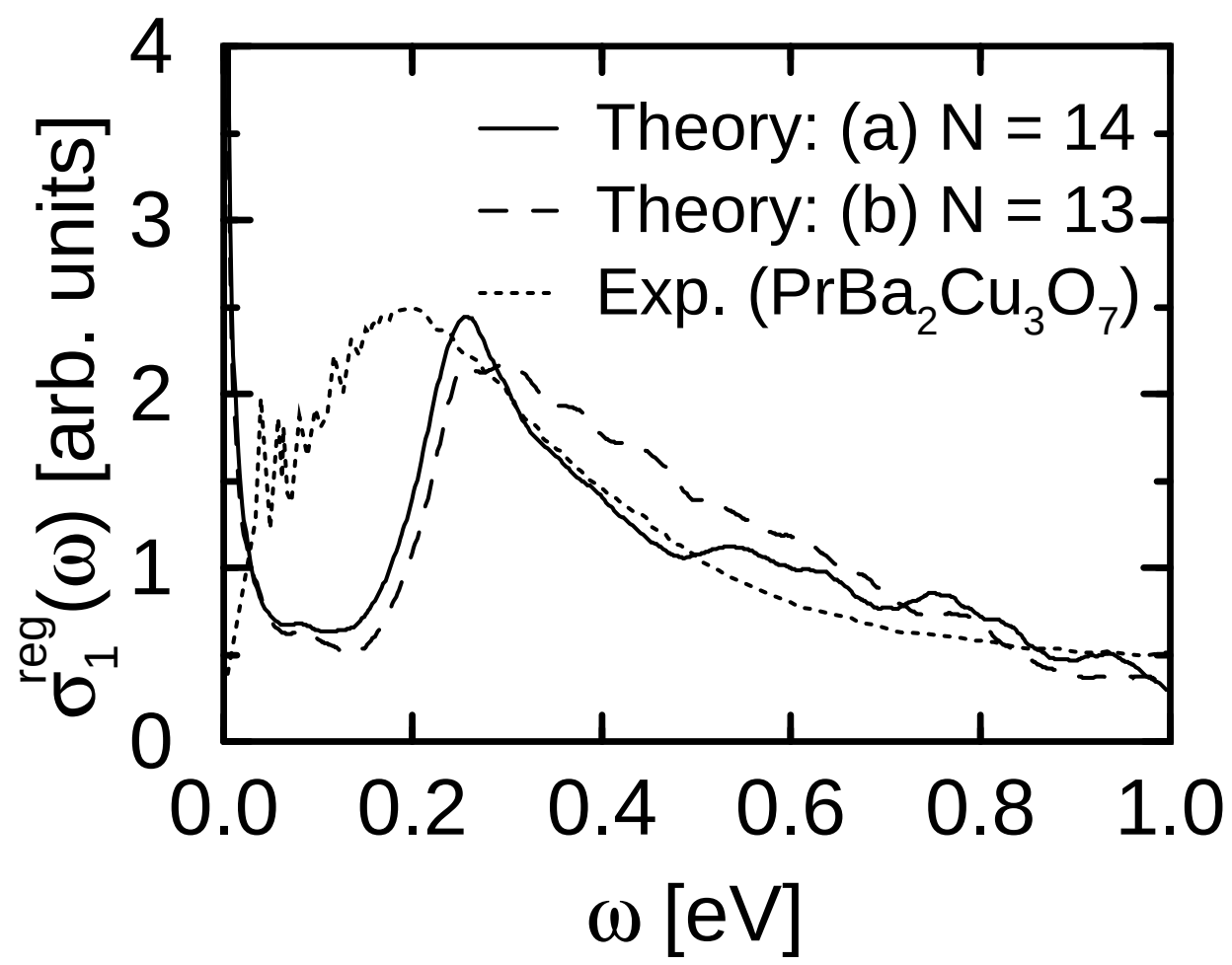

**Figure** 15: A comparison of the experimental spectrum obtained in Ref. [1] with two curves calculated using OBC and parameters (a) $N$ = 14, $E_B$ = 0.65, and (b) $N$ = 13, $E_B$ = 0.55, both at $n_e$ = 6, Ω = 0.2.

There are three possible explanations for the shortcoming at low frequencies: (i) Since in our model, the position of the main peak, and therefore the size of the quasi-gap is determined by the energy of the lowest particle-hole excitation of the corresponding non-interacting system, it is clear, that it should diminish as a function of increasing system size. Hence, it is plausible to assume, that the contribution of longer chains ($N > 14$) to the total spectrum in the real samples would fill in the missing spectral weight at low energy. Such long chain segments should certainly be present in these materials, provided that the oxygen vacancies occupy random positions, and do not form an ordered superlattice, for which there is no experimental evidence. Unfortunately, the limitations posed by the available computer power, make it impossible to demonstrate this explicitly by a calculation. However, the comparison in Fig. 16 between the spectra for increasing system size clearly shows the shift of the peak.

(ii) The second explanation could be that the approximation of Ansatz II slightly underestimates the contribution of renormalized ($\omega < \omega_{\text{peak}}$) versus unrenormalized (weight of main peak) charge motion, so that in reality (using the full phonon Hilbert space) there would not be a quasi-gap, but rather a smooth decay of the conductivity as we approach $\omega = 0$. If this was the case, we should see at least the onset of such a shift of spectral weight, when comparing the results from Ansatz II with the calculations performed using the two additional excited phonon states $|1\rangle_l, |\tilde{1}\rangle_l$.

Fig. 17 shows a comparison between the two approaches for a 9-site lattice with $n_e$ = 4, Ω =

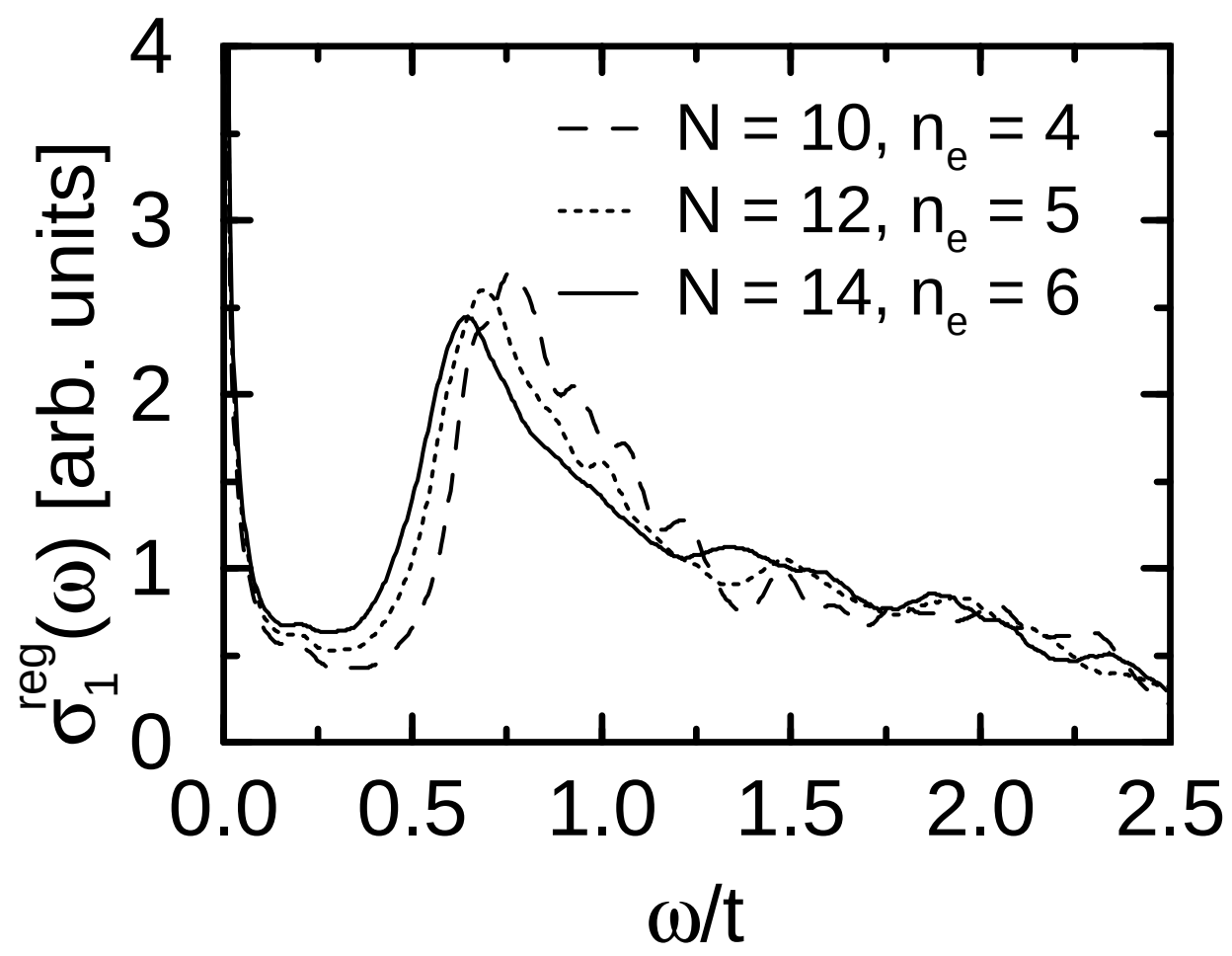

**Figure** 16: A comparison of spectra $\sigma_1^{\text{reg}}(\omega)$ obtained for different combinations of lattice sizes and fillings $[N, n_e] = ([10,4],[12,5],[14,6])$ using OBC and fixed values of $E_B = 0.65$, $\Omega = 0.2$.

0.2, and different coupling strengths. The e-ph coupling was chosen such that the qualitative shape of the peak and the high-frequency tail are similar for the two approximations. We observe an excellent agreement between the two cases over the entire frequency range, *i. e.*, the low-frequency weight is *not affected* by the reduction of the phonon Hilbert space. This agreement supports the hypothesis[25] that the neglected phonon states in Ansatz II can be taken into account by a renormalized e-ph coupling. In other words, the spectrum we obtain from a calculation using Ansatz II at coupling $E_B$ presents an excellent approximation of the true spectrum (using the full Hilbert space) at a coupling strength $E'_B > E_B$. At least this should be the case if both, the particle density $n$, and the phonon frequency $\Omega$ are not too small. As we showed in Ref.[25], at small density, and $\Omega/t \leq 0.1$, excitations on the scale $\Omega$ become relevant, and this leads to a qualitative difference between the results from Ansatz II and those from the true Hamiltonian. Note, also, that the additional phonon states lead to a slight smoothening of the spectra as compared to Ansatz II.

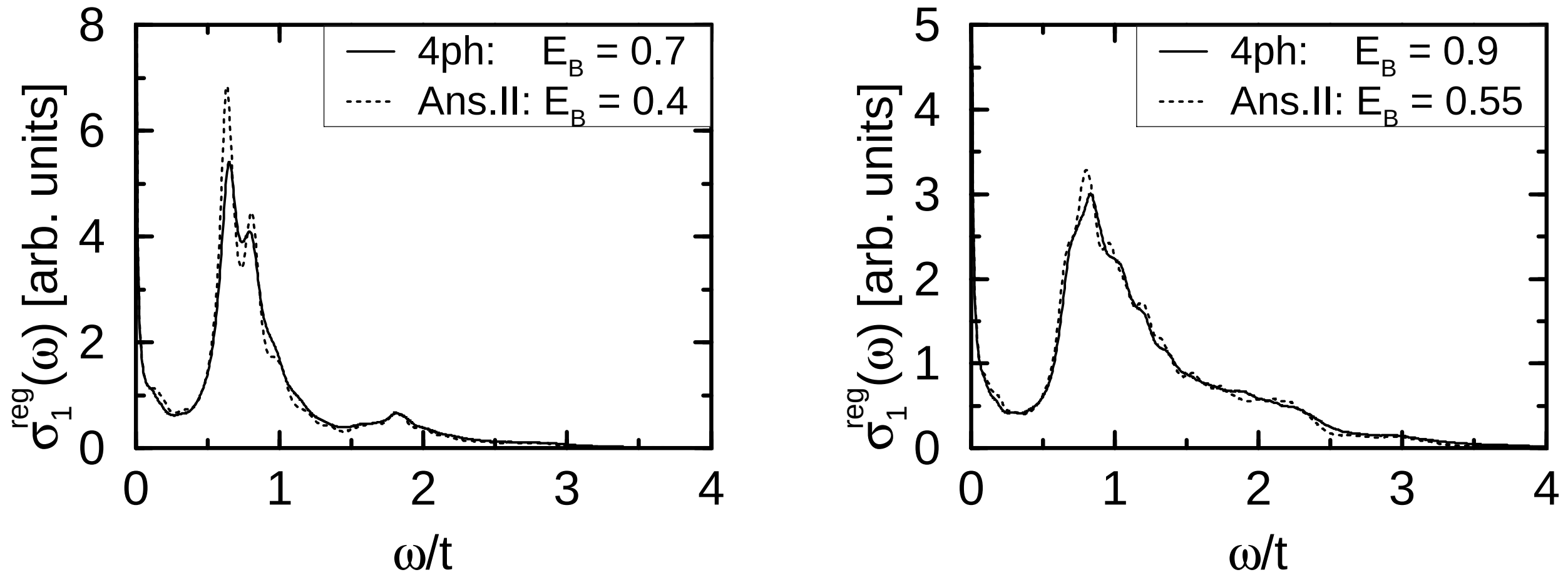

**Figure** 17: A comparison of $\sigma_1^{\text{reg}}(\omega)$ obtained from Ansatz II with the results using a four-phonon basis for different couplings. The spectra were obtained for $N = 9$ with OBC at $n_e = 4$ and $\Omega = 0.2$.

(iii) Finally, we should consider the possibility that the scattering from spin exchange could be enhanced by the e-ph coupling as compared to the simple *t*-*J* model (see the discussion in section II). Naively, one expects a competition between the spin and the polaronic e-ph coupling. The former favours singlet bonds between electrons on neighbouring sites, *i. e.*, an effective attraction, whereas the latter leads to an effective nearest neighbour repulsion, *i. e.*, CDW correlations. This competition might lead to interesting effects, which we investigate in more detail in Ref. [25].

In Fig. 18 we show the spectra calculated for a 12-site lattice with $n_e = 5$, total spin $z$-component of $S_z = 1/2$, parameters $E_B = 0.65, \Omega = 0.2$, and different values of the spin exchange coupling $J$. The ground state for all the parameters has the lowest possible value for the total spin, $S = 1/2$, as found in the simple *t*-*J* model. We note that the spin degrees of freedom enhance the low-frequency weight below the main peak, and shift the latter to lower energy, while the high-frequency weight is reduced. The additional low-frequency weight increases as a function of $J$. This effect of the spin exchange on the low-frequency region is very similar to what happens when the e-ph coupling strength is reduced. A possible explanation for this behaviour could therefore be that the nearest neighbour spin exchange coupling, which favours electrons on neighbouring sites, leads to an effective weakening of the repulsive CDW-correlations on the scale $\omega \lesssim J$. This can be verified in the inset of Fig. 18, where one observes a strong similarity between a spectrum at $J \neq 0$ and some $E_B$, with a spectrum at $J = 0$ and a smaller $E'_B < E_B$. Apart from this effect, the spin degrees of freedom do not play a significant role.

From these considerations, it seems most likely that the missing weight should come from the contributions of larger chains. We expect that this should fill in the quasi-gap. Even though we cannot explicitly calculate a spectrum for this situation, our results strongly suggest that a spectra obtained from a superposition of random length chain segments (similar to the case we calculated in section III without e-ph coupling) would give good agreement with the experimental one.

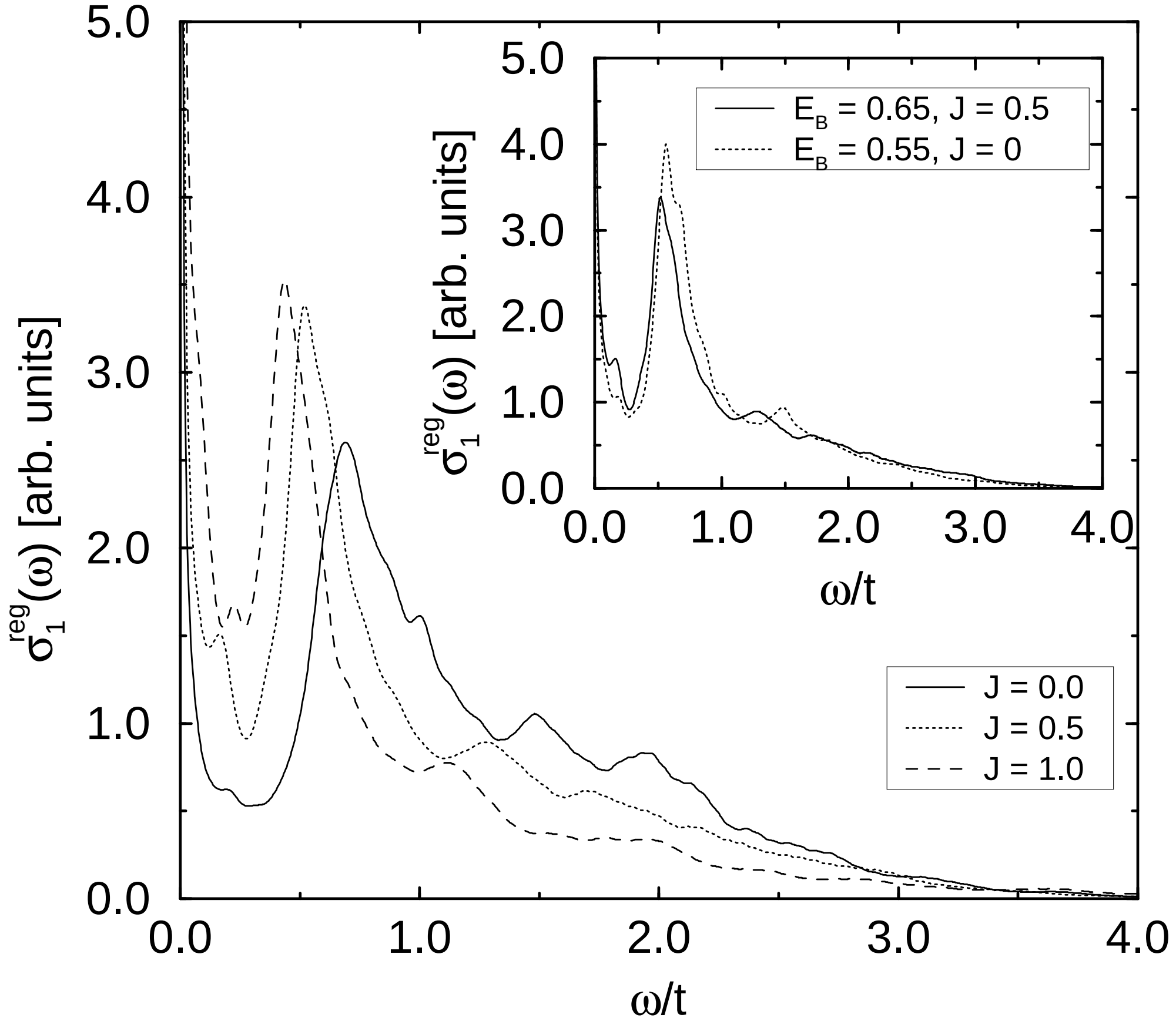

**Figure** 18: A comparison of $\sigma_1^{\text{reg}}(\omega)$ obtained from Ansatz II at different values of the spin exchange coupling $J = 0.0, 0.5, 1.0$, and fixed $N = 12, n_e = 5, E_B = 0.65, \Omega = 0.2, S_z = 1/2$ using OBC. The inset contrasts the case of $J = 0.5$ with a calculation for $J = 0, E_B = 0.55$.

## VII. Conclusions

In this article, we discussed different physical mechanisms which could be responsible for the experimentally observed frequency dependence of the optical conductivity in the $CuO_3$-chains of the 1-2-3 compounds. The main features of the conductivity are a broad mid-infrared peak at $\omega \approx 0.2$eV, and a subsequent slowly falling (much slower than $1/\omega^2$) high-frequency tail.

In view of the similarity between the CuO bonding patterns of the $CuO_2$-planes, and the $CuO_3$-chains, we argued that a 1D $t$-$J$-model, derived from a multi-band Hubbard model should capture the relevant low-energy physics of the chains. However, the known fact that this model does not show any substantial optical absorption (due to the charge-spin decoupling in 1D), forced us to consider additional terms in the Hamiltonian, which could lead to the observed strong scattering, and the resulting absorption.

The inherent tendency of oxides towards the formation of defects naturally suggests to include the resulting perturbations of the charge motion in a model of the electronic structure. In the case of the 1-2-3 compounds, the oxygen vacancies on the O(1)-site connecting two neighbouring Cu(1) in the chains present the most serious and also most common type of defect.

The simplest way to model the vacancy disorder is to assume that a vacancy acts as an almost impenetrable barrier. This leads to a decomposition of an individual chain into effectively disconnected segments, between which charge motion is severely hindered. This effect can be incorporated in the Hamiltonian by a strongly reduced hopping matrix element between the lattice sites next to a vacancy. For the optical conductivity, this has the consequence that the dc conductivity vanishes, and its weight is transferred to finite frequencies.

Neglecting the magnetic term in the $t$-$J$-model, we calculated the optical conductivity for such a model, assuming a uniform random distribution of vacancies, characterized by the mean length of an individual segment. Fixing the energy scale by setting $t = 0.4$eV, we found that this model exhibits a low-energy peak, but it is neither able to correctly reproduce the peak position nor does it show the observed high-frequency tail. For a more realistic model, where in addition to the disorder in $t$, we also included a repulsive potential on the sites neighbouring the vacancy, we did not find any better agreement with the experimental spectrum.

As a second possibility, we included *potential* disorder, which can be modeled by random on-site energies on each lattice site. The ac conductivity of such models is well known to show (i) a low-energy peak, the position of which depends on the disorder strength, and (ii) a high-energy tail. Therefore it appears to be a promising candidate to explain the experiment. Indeed, we found that it is possible to fit the measured spectrum very accurately by this model. However, a quantitative analysis revealed that a very large disorder strength ($W \approx 1.2-1.6$eV) is needed to obtain the correct peak position. There is no experimental evidence for such strong disorder in the real materials, so it remains unclear whether this is the correct model. A possible type of defect, which could in principle produce such strong potentials, is an interstitial oxygen atom between two Cu(1) belonging to neighbouring chains. But the density of such defects ($\approx 10\%$) required to produce such a strong random potential on *each* site seems unphysically large.

As a third likely source of strong scattering in the chains we considered polaronic electron-phonon coupling. In particular, we argued that the phonon mode which involves the displacement of the apical oxygen (O(4) site) in **c**-direction, should couple rather strongly to the holes in the chains. We showed that this coupling is well described by an additional Holstein term in the Hamiltonian, the local ion displacement being linearly coupled to the local charge density. For simplicity, we assumed this phonon mode to be dispersion-less. Again neglecting the spin degrees of freedom, the effective model becomes the standard Holstein molecular-crystal model for spinless fermions.

In consideration of the importance of the effective electron-electron interactions, we chose a numerical technique for the calculation of $\sigma(\omega)$, which is capable of dealing with such interactions in a clean way, the Lanczos-method. Apart from the limitation to finite lattices, the drawback of this method when applied to an e-ph Hamiltonian is, that one is forced to restrict the possible phonon states to some finite subspace, in order to obtain a finite-dimensional total Hilbert space.

In an accompanying paper [25], we investigated different possible variational subspaces for the phonons, each of them allowing two states at a given lattice site. The best choice for the two local phonon states, which we called Ansatz II, was found to be the vacuum states of the undisplaced, and the displaced oscillator $|0\rangle_l, |\widetilde{0}\rangle_l$. The agreement from Ansatz II of the parameters entering the effective Hamiltonian with the exact ones, as well as with the QMC data was excellent.

For a comparison with the experimental spectrum, we used Ansatz II in the presence of open boundary conditions, which is the simplest way to include the effect of the oxygen vacancies. We found that the effective electron correlations lead to the appearance of the high-frequency tail also in this situation, provided the density is close to 1/2, which represents the estimated experimental value. In the low-frequency region, we also found the characteristic peak, the origin of which can be traced back to the lowest particle-hole excitation of the non-interacting system. At even lower energies, the absorption due to the renormalized motion of the domain walls partly closes the quasi-gap.

When faced with the experimental spectrum, the agreement is fairly good. While the high-frequency tail is reproduced very well, at low-energy there is some discrepancy: The conductivity falls off two sharply on the low-energy side of the main peak, and the quasi-gap leaves some missing weight. Most likely, this shortcoming is due to the fact that only chains with length $N \leq 14$ could be considered in our calculations. The contribution of longer chains has to be expected to close the quasi-gap, since its size diminishes for increasing chain lengths. Adding the spin degrees of freedom was shown to lead to additional absorption in the low-frequency region, and a slight shift of the main peak to lower energies, without strongly affecting the high-frequency tail. However, for physical values of $J/t \lesssim 0.5$, the overall influence of the spin degrees of freedom was found not to be very significant, as expected due to charge-spin separation. The inclusion of additional phonon states mainly has the effect of renormalizing the e-ph coupling strength to larger values. Apart from a slight smoothening of the spectra, it does not alter the results.

In conclusion, there are two likely mechanisms which could give rise to the observed conductivity of the $CuO_3$-chains in the 1-2-3 materials: While the first one (strong potential disorder) gives excellent agreement with the experimental spectrum it suffers from the difficult justification of the required disorder strength. The proposed polaronic e-ph coupling can be well justified assuming reasonable physical parameters, and also gives fair agreement with the experiment.

It would be highly desirable to have an experimental verification for one or the other mechanism. The total spectral weight expressed in the effective carrier number per unit cell was found to be very similar in both models $n_{\mathrm{eff}} \approx 0.4 - 0.45$, and compared well with experiment $n_{\mathrm{eff}} \approx 0.5$. One possibility to distinguish the two mechanisms could be the investigation of the 'isotope effect' on the conductivity spectrum. If the e-ph model is applicable, the substitution of the apical oxygen from $O^{16}$ to $O^{18}$ should lead to a reduced phonon frequency $\Omega$, and hence to an increased renormalization factor in which $\Omega$ enters exponentially. In this case a shift of the main peak to higher frequency would be expected. Furthermore, one should observe a sign of short-range CDW order resulting from the polaron-polaron interactions in this case.

**Acknowledgements**

The author would like to thank Prof. T.M. Rice for suggesting this problem to him, for many helpful and clarifying discussions, and his continous support. Many useful conversations and comments by H. Fukuyama, E. Heeb, R. Hlubina, D. Poilblanc, P. Prelovšek, H.B. Schüttler, M. Troyer, H. Tsunetsugu, and F.C. Zhang are gratefully acknowledged. I would also like to thank G. Blatter for his continous support. This work was supported partially by the Swiss National Science Foundation.